\documentclass[acmsmall,nonacm,screen,review = False]{acmart}

\renewcommand\footnotetextcopyrightpermission[1]{} 
\usepackage{graphicx}
\usepackage{subcaption}
\usepackage{makecell}
\usepackage{tabularx}
\usepackage{amsmath}
\usepackage{multirow}
\usepackage{paralist}
\usepackage[T1]{fontenc}
\usepackage[autostyle]{csquotes}
\AtBeginDocument{%
  \providecommand\BibTeX{{%
    \normalfont B\kern-0.5em{\scshape i\kern-0.25em b}\kern-0.8em\TeX}}}

\setcopyright{acmcopyright}
\copyrightyear{2018}
\acmYear{2018}
\acmDOI{10.1145/1122445.1122456}

\acmConference[Woodstock '18]{Woodstock '18: ACM Symposium on Neural
  Gaze Detection}{June 03--05, 2018}{Woodstock, NY}
\acmBooktitle{Woodstock '18: ACM Symposium on Neural Gaze Detection,
  June 03--05, 2018, Woodstock, NY}
\acmPrice{15.00}
\acmISBN{978-1-4503-XXXX-X/18/06}

\newcommand{\cmttfont}[1]{\fontfamily{cmtt}\selectfont}
\DeclareTextFontCommand{\textcmttfont}{\cmttfont}

\begin{document}

\title[Dangerous Speech on Twitter in India]{Insights Into Incitement:  A Computational Perspective on Dangerous Speech on Twitter in India}

\author{Saloni Dash}
\affiliation{%
 \institution{Microsoft Research}
 \streetaddress{Lavelle Road}
 \city{Bengaluru}
 \state{Karnataka}
 \country{India}}

\author{Rynaa Grover}
\affiliation{%
 \institution{Microsoft Research}
 \streetaddress{Lavelle Road}
 \city{Bengaluru}
 \state{Karnataka}
 \country{India}}

\author{Gazal Shekhawat}
\affiliation{%
 \institution{Microsoft Research}
 \streetaddress{Lavelle Road}
 \city{Bengaluru}
 \state{Karnataka}
 \country{India}}
 
 \author{Sukhnidh Kaur}
\affiliation{%
 \institution{Microsoft Research}
 \streetaddress{Lavelle Road}
 \city{Bengaluru}
 \state{Karnataka}
 \country{India}}
 
\author{Dibyendu Mishra}
\affiliation{%
 \institution{Microsoft Research}
 \streetaddress{Lavelle Road}
 \city{Bengaluru}
 \state{Karnataka}
 \country{India}}

\author{Joyojeet Pal}
\affiliation{%
 \institution{Microsoft Research}
 \streetaddress{Lavelle Road}
 \city{Bengaluru}
 \state{Karnataka}
 \country{India}}

\renewcommand{\shortauthors}{Dash et al.}

\begin{abstract}
Dangerous speech on social media platforms can be framed as blatantly inflammatory, or be couched in innuendo. It is also centrally tied to who engages it -- it can be driven by openly sectarian social media accounts, or through subtle nudges by influential accounts, allowing for complex means of reinforcing vilification of marginalized groups, an increasingly significant problem in the media environment in the Global South. We identify dangerous speech by influential accounts on Twitter in India around three key events, examining both the language and networks of messaging that condones or actively promotes violence against vulnerable groups. We characterize dangerous speech users by assigning  \textit{Danger Amplification Belief} scores and show that dangerous users are more active on Twitter as compared to other users as well as most influential in the network, in terms of a larger following as well as volume of verified accounts. We find that dangerous users have a more polarized viewership, suggesting that their audience is more susceptible to incitement. Using a mix of network centrality measures and qualitative analysis, we find that most dangerous accounts tend to either be in mass media related occupations or allied with low-ranking, right-leaning politicians, and act as ``broadcasters'' in the network, where they are best positioned to spearhead the rapid dissemination of dangerous speech across the platform. 
\end{abstract}




\maketitle

\section{Introduction}
Incitement and polarization are increasingly central to political environments around the world, and there are arguably few better examples of ``technology for social bad'' than the unintended consequences of social media use that have enabled such divisions between people.  The increasing use of social media and micro-blogging sites as a source of news and social engagement has enabled political parties to directly engage with, and often radicalize their vote base in political systems throughout the world. We see this particularly in countries like India, where politicians, instead of reaching out to their constituents through the mainstream media, communicate unhindered through dedicated, curated Twitter feeds. This has resulted in what may seem at first glance to be a more participatory democracy where citizens engage directly with their representatives, but in practice it enables the misuse of platforms for spreading propaganda, hate and misinformation \cite{mahapatra2019polarisation}.  

The affordances of social media enable affective virality, where people engage with content not for its presumed veracity, but rather for the emotional content -- indeed, for much of the last decade, social media has enabled various forms of violence and vigilantism sparked by misinformation. In India, the term \textit{WhatsAapp Lynchings} has been used to refer to violence triggered by social media discourse overlaid on existing antagonisms \cite{banaji2019whatsapp}. Various arguments have proposed that social media enables the articulation of several kinds of prejudice that predate technology, and rather than a cause, they should be seen as explainers of what trigger violent behavior \cite{chinmayi2019}. 


Inflammatory speech on social media can have deleterious effects while simultaneously offering legal deniability to instigators who can choose between direct provocation and masked innuendo. Thus, social media platforms find themselves at the core of these debates -- balancing the legal ramifications of such speech on their platforms with the economics of free speech on usage, which more often than not, benefit from incitement \cite{rathje2021out}. The goals of nation states and policy regimes to regulate platforms must also be examined through the lens of their own attitudes towards free speech and maintenance of power. The ongoing tensions between the Government of India and social media platforms offer a case in point \cite{oxfordtwitter}. While the government has been asking for regulations which are already in place in many other nations, its attempts to shut down online dissent forebode poorly on its motives. 

Within the context of these contradictions, we examine the spread of \textit{dangerous speech} on Twitter in India.
Dangerous Speech is \textit{``any form of expression (e.g. speech, text, or images) that can increase the risk that its audience will condone or participate in violence against members of another group''} \cite{speechdangerous}. Attributes like the influence of the speaker, the susceptibility of the audience, the inflammatory nature of the speech, its medium of dissemination and context, characterize dangerous speech. In India, these conditions have materialised not only in its post-partition riots and inter-religious tensions, but concerns of the susceptibility of social media users to disinformation and dangerous speech, have become centre-stage in determining whether individuals in power are to ``blame'' for the aftermath of their incitement. These observations have informed our dataset, which is a collection of individuals who are influential in the Indian setting. The dataset comprises of 6k ``influencers'', individuals who derive their legitimacy either solely or partly through online operation, or have an offline sphere of expertise such as entertainers, journalists \cite{dash2021divided} and 26k politicians from from two major national parties in India -- the national incumbent Bharatiya Janata Party (BJP) and the opposition party Indian National Congress (INC). 

\begin{figure}[!htb]
    \centering
    \includegraphics[width = 0.8\linewidth]{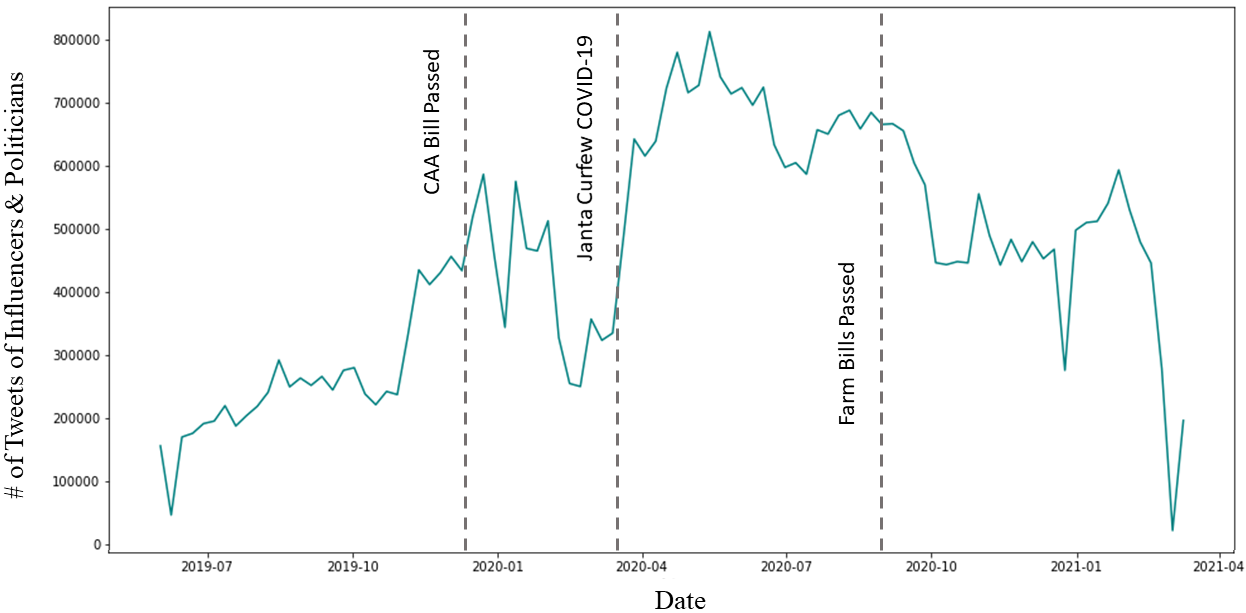}
    \caption{\textbf{Tweeting Activity of Influencers \& Politicians between June 2019 and April 2021}}
    \label{event_timeline}
\end{figure}

Specifically, we consider three polarizing issues that have seen spikes in online traffic from influential individuals in India from 2019-2021, as seen in Figure \ref{event_timeline}.
In the order of their onset, they are the Citizenship Amendment Act (CAA), the Tablighi Jamaat congregation during the COVID-19 pandemic in India and the farmers' protests in response to the the Indian Farm Bills Act of 2020.

The CAA, which creates a legal framework to give refugees access to Indian citizenship, does so based on religious persuasion, excluding Muslims from neighboring countries. The act was the first time that religion-based exclusion was proposed as part of access to citizenship in India, and was to be enacted alongside a Nationwide Register of Citizens (NRC) in order to identify ``illegal immigrants'' living in India by requiring fairly detailed historical documentation of residency. This stoked fears of losing citizenship among India’s Muslims as they would have to prove they are not from a neighboring nation, and led to widespread protests all around the country.

The Tablighi Jamaat incident refers to a religious conference of members of the Muslim sect that was held in New Delhi's Nizamuddin Markaz Mosque. The gathering that began in early March 2020, had thousands of visitors who had only partially dispersed by the time a nationwide lockdown was announced later in the month. Thousands of COVID-19 cases were attributed to the gathering and it was consequently termed a superspreader event. While the scale of the gathering's effects is still disputed, as well as whether the organizers and attendees were in willful violation of the law at the time, the event nonetheless triggered a media frenzy. A significant part of this took place on social media, with the viral hashtag \textcmttfont{\#CoronaJihad}, driving a discourse that the Muslim community was intentionally spreading the virus throughout the country. 

The Indian Farm Bills passed in parliament in September 2020, were part of a subsidy reform initiative that allowed the entry of corporates into certain domains of crop trading, which were formerly protected. This led to protests by farmers, particularly in states of North India where the new configurations of crops and purchase prices brought threats to existing financial equations and livelihoods. The social media backlash against the protests focused particularly on Sikhs, who formed a large component of the protesting farmers, and presented them as acting antithetically to the interests of the broader nation.

Across all of the events, we find that the dangerous speech framework is well suited to make sense of the behavior we observe on Twitter, for both content and network characteristics. Messages do not spread evenly but are instead dependent on the network of the audiences and their interactions with tweets. Polarizing issues like these, often co-occur with forms of homophily in shaping sides and gain momentum from inflammatory messaging, something that research on affective engagement has repeatedly shown. Attributes of dangerous speech also play a role in creating or reinforcing influence among key nodes in the Indian Twitter network –- thus, while on one hand social media influencers amplify dangerous speech, they also benefit from it through increased following and engagement. 

Despite its relevance, dangerous speech has not been as widely studied as other forms of harmful speech, like hate speech, particularly in the context of the Global South. The nuances of the framework make it tougher to analyse in a computational setting. This paper is a first step towards addressing that. We investigate the following research questions -- \\
1) \textit{Can the conceptual framework of dangerous speech be adopted in a computational setting?}\\
2) \textit{Are there any distinguishing characteristics of dangerous speech users in comparison to other users?}\\
3) \textit{What were the predominant antagonistic narratives in the dangerous speech messaging and whether they were foreboding of the violence that occurred during the periods?}\\
Overall, we make several significant contributions, through which we not only successfully quantify the dangerous speech framework but also characterize dangerous speech in India -- i) we identify dangerous speech by using a lexicon based subsampling method followed by manual annotation ii) we compute a \textit{Danger Amplification Belief} (DAB) score, for each of the users by incorporating the influence of the individual in the retweet network iii) we further characterize the dangerous users in terms of account attributes like tweeting activity, following etc. and derived attributes like polarity, which measure the susceptibility of the audience.
iv) we identify the dominant and supplementary narratives that accompany dangerous speech through these periods, their implications for target communities as well as the applicability of our findings in other contexts.

The remainder of the paper is structured as follows. Section \ref{indian_context} offers a case for the use of the dangerous speech framework, and highlights its relevance within the Indian context. The Indian context also has broader implications for information environments that are comparable elsewhere in the world where people are newly online, and where media environments come bundled with risks of polarization or manipulation. Section Section \ref{related_work} notes past computational work in the area, traces the conceptual framework of dangerous speech, and discusses the role of influential speakers within social media information networks and the political context within which we situate our study. Section \ref{method} details the pipeline for quantifying dangerous speech as well as computing derived user attributes like polarity. \ref{ds_character} characterizes the dangerous users and further highlights the differences in comparison to hateful users. Section \ref{ds_content} delves into the dominant dangerous speech narratives in each of the three events, validates the characteristics of dangerous users with a ground-truth analysis of the users within our dataset, and extracts general patterns of dangerous speech messaging across all the events and their applicability to other contexts. In conclusion, section \ref{discussion} reflects on the consequences of the findings, the limitations of our study and potential directions for future work in Global South settings.

\section{Dangerous Speech in the Indian Context} \label{indian_context}
Our adaptation of the dangerous speech framework for studying inflammatory speech on Indian Twitter, over the use of the popularly studied term, hate speech, is motivated by several factors. Firstly, the lack of a commonly accepted definition of hate speech can prove to be counterproductive in the Indian setting.
As previously noted, when politicised, the label of hate speech can be used to further institutional agendas and suppress free speech \cite{chinmayi2019, pohjonen2017extreme}. Because these complications have increasingly arisen in the Indian state's relationship with social media companies, for both agenda amplification and blame shifting \cite{narrain2017dangerous}, we find that the dangerous speech framework is valuable in providing a critical look at the role of powerful actors in furthering incitement. 

Moreover, the dangerous speech framework goes beyond the binary question of whether a text is hateful or not, to account for the legitimacy and power held by speakers, which can alter capacity of the speech to cause harm. The emphasis on looking beyond the egregiousness of the message is particularly relevant in the Indian context, as we show further below, where seemingly innocuous tweets from an influential speaker can trigger violence in a susceptible audience.

\begin{table}[!htb]
\centering
\begin{tabular}{|p{0.7\textwidth}|p{0.2\textwidth}|}
\hline
\textbf{Tweet}  & \textbf{Category}       \\ \hline

\textit{chilling conversations on delhi streets where sympathy for farmers is giving way to antisikh prejudice fuelled by false media narratives connecting farmer protests to terrorists and khalistanis.}  & NDS            \\
\hline
\textit{thousands of packets of free biryani, rs 500-700 reportedly spent on each protester \& other event expenses? who is sponsoring the shaheenbagh protests? \#shaheenbaghtruth \#shaheenbaghcracks.}  & DS, but not HS \\
\hline
\textit{with the growing islamic violence hate and the venom they're spitting to spread coronavirus  across the country another rama another krishna another parashurama must take birth to defeat these demonic beasts and to prevail the peace \#nizamuddinmarkaj \#coronajihad} & HS and DS      \\ \hline

\end{tabular}
\vspace{6pt}
\caption{\textbf{Instances of Dangerous Speech (DS), Hate Speech (HS) and Not Dangerous Speech (NDS)}}
\vspace{-14pt}
\label{ds_eg}
\end{table}

In order to further highlight the contextual nuances, often requiring human interpretation, that the dangerous speech framework captures, we discuss the varying usages of speech in tweets in Table \ref{ds_eg}. The first tweet, while using \textcmttfont{terrorists} or \textcmttfont{khalistanis}, does so to lament the state of affairs, and therefore is not eligible to be classified as dangerous speech. The third tweet is antithetical in tone to the first one, and uses dehumanising language (\textcmttfont{demonic beasts}, \textcmttfont{venom they're spitting}), to call for the reincarnation of a Hindu deity to ``defeat'' the target group of Muslims. This language is dangerous and can explicitly be classified as hate speech. 
However, the second tweet, which makes no mention of commonly used slurs or abuses, would not necessarily be classified as hate speech but has several attributes of dangerous speech. On the surface, the tweet can seem like public-interest questioning, yet its insinuations frame the CAA/NRC protesters as paid and greedy, effectively chipping away at the legitimacy and ``true intentions'' behind the sit-in protests led by Muslim women. 

\begin{figure}[!htb]
    \centering
    \includegraphics[width = 0.8\linewidth]{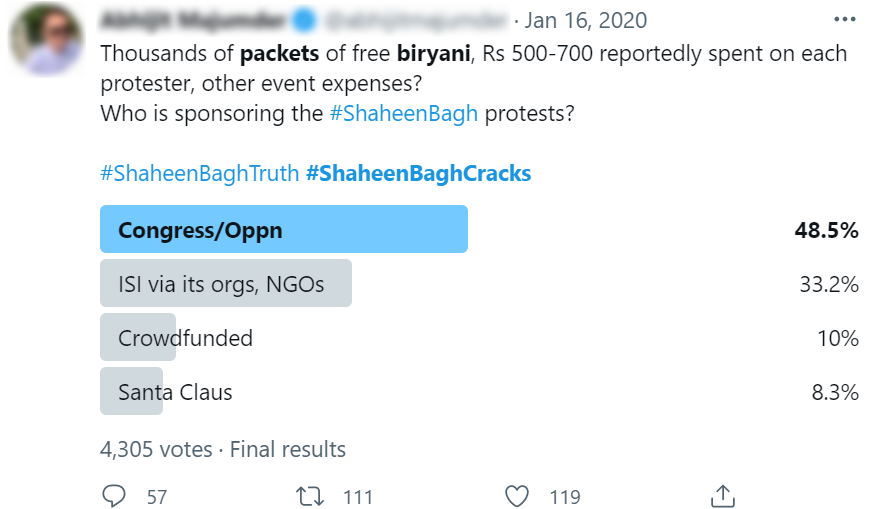}
    \caption{\textbf{Instance of Dangerous Speech on Twitter Published by an Influential Journalist}}
    \label{biryani_tweet}
\end{figure}

As mentioned previously, the influence of the speaker is a significant distinguishing feature of dangerous speech. In Figure \ref{biryani_tweet}, we find that the second tweet from Table \ref{ds_eg} was published by an influential journalist, who has approximately 500k followers, including India's Prime Minister Narendra Modi. The tweet is in the form of a poll, inviting interactions, with options that the protesters are being funded by -- the opposition, Pakistani intelligence agencies, crowdfunding and ``Santa Claus''. The framing is biased and assumes a priori that the protests are disingenuous. Most respondents choose the first two categories, which attributes malintent to vested interests. Because it is framed as a message that invites and builds on public opinion, the tweet purports to present a democratized view of what is happening on the ground. Overall, this tweet enjoyed the most retweets, comments and engagements by a significant margin compared to the other two, allowing it to cement conspiratorial allegations. That the speaker is a journalist, verified and followed by prominent politicians, gives this messaging credibility, and speaks of the dangerous environment it can reinforce. A month after the tweet was posted (and many others like it, in a period of heightened online dangerous speech), Muslims were disproportionately killed in the Delhi Riots in February 2020. Therefore, what is collectively interpreted to be rational violence, and who deserves justice can be skewed through speech acts, where some victims are presented as more ``legitimate'' than others.

\section{Related Work} \label{related_work}
In this section, we examine previous computational approaches adopted for the detection of speech considered to be harmful, such as hate speech and fear speech, through distinct conceptualisations and contrast them with definitional elements of the dangerous speech framework. Additionally, we discuss the role of influential users in propagating dangerous speech and contextualize our study within India’s political landscape.

\vspace{-3pt}
\subsection{Computational Approaches for Speech and Harm
}


A rise in public awareness over the past decade has paralleled the media and scholarly attention given to hate speech within online social networks. The use of computational approaches towards the study of this phenomena can be an effective means to understand the nature of socialised ire in different societies and reflect on pathways to reduce harm. In the Indian context, several deep learning approaches, like Hierarchical LSTMs \cite{santosh2019hate}, character-level CNNs \cite{bohra2018dataset} and transformer models like XLNET \cite{banerjee2020comparison} have been successfully applied to Hindi-English code-mixed content for hate speech detection. However as the study of hate speech matures into a sub-discipline, scholars have also noted the shortcomings of existing research on the theme. The most glaring among them is the overrepresentation of case studies about the political and sociological environments of the USA in the overall research on the theme \cite{udupa2020hate, tontodimamma2021thirty}. As \citet{pohjonen2017extreme} note, a largely western understanding of hate speech can be “superimposed” on regions far away, even as there are contesting and contextual practices that emerge in a communities’ view of what is considered harmful or unacceptable forms of speech.

Additionally, \citet{saha2021short} have observed that a binarized definition of hate speech when based merely on the content, can set higher thresholds for what texts raise alarm, and ignore other harmful narratives that flow unbridled across communication environments. Reflective of the distinct challenges posed by such speech, the authors have developed a model to identify``fear speech” on Indian WhatsApp groups \cite{saha2021short}. The authors’ understanding of fear speech builds on Buyse’s \cite{buyse2014words} conceptualisation, which aims to describe the forms of speech that cross the legal boundaries of free speech. In capturing what the authors understand to be high as well as low toxicity speech acts, their motivations run conceptually parallel to our adaptation of the dangerous speech framework. However, the study, with its significant findings on the use of politicised propaganda against Muslims, is based on Whatsapp, while the platform of our research, Twitter, introduces distinctive network characteristics of influence, source and audiences, compared to more intimately linked messaging apps. 

In general, large scale quantitative techniques divorced from regional nuances are cursory at best and deleterious at worst  \cite{bishop2018anxiety,sap2019risk}. Our work around dangerous speech attempts to be mindful of ground-level realities by tracing temporal and community level gradients through which target groups in India have been characterized. Although it takes important cues from both, our proposed framework differentiates dangerous speech from hate speech, which finds usage in a broader variety of contexts and focuses on content rather than speaker influence and intent \cite{udupa2020hate}, as well as fear speech, which intends on instilling fear as a primary driver of polarisation \cite{saha2021short}.

\subsection{Elementary Characteristics of Dangerous Speech}
To justify the use of our proposed framework, we delineate online dangerous speech by noting its distinguishing conceptual elements. When it comes to the popular term describing harm within speech acts, the term ``hate speech” is often employed to shed light on the egregiousness of the content of such speech, whereas dangerous speech assigns a substantial weight to the sources of the content. This is because the emergence of the dangerous speech framework is tied to historical reflection about the power of speech to incite violence, long before the rise of social media \cite{benesch2012dangerous}, and the authority afforded to its instigators.   

\citet{saha2021short} have observed that the adaptation of the framework for computational studies can be challenging as it is difficult to trace a direct link between dangerous speech and violent events. However, as recent definitions \cite{speechdangerous} have noted, speech in this realm is dangerous not only in its ability to cause violence, but for the violence to be condoned, and therefore, normalised. Considering the emergence of vigilante violence as well as riots attributed to social media in India, we consider this conceptualisation as it accommodates for material harms of dangerous speech in the lived experiences of people targeted by it, a reality that broad-based quantitative studies risk minimising. A recent paper by \citet{alshehri2020understanding} also centres around building and detecting a dataset around dangerous speech on Arabic Twitter. However, the authors use an original definition of dangerous speech and do not refer to the framework conceptualised by Benesch \cite{speechdangerous} across qualitative studies and public engagements. Therefore, ours is the first attempt to quantify the dangerous speech framework, by \citet{benesch2012dangerous} in a quantitative setting. 

Qualitative studies on the dangerous speech framework provide valuable insights into the strategies employed among those who exercise it. Instigators have strategically and systematically dehumanised target groups \cite{speechdangerous} by using coded signals for legibility within their in-group \cite{speechdangerous}. This is similar to dog-whistling, for dangerous speech can be attached under social and political arguments, to convert dangerous ideology into permissible online text \cite{caiani2021online}. These tactics can also reflect those of extremism and radicalisation. Here, instigators strategically invoke the in-group’s loyalty, causing shifts in collective morality and permissibility \cite{giner2012dehumanization}. At later stages, this displaced sense of justice can provide an impetus for calls about the target group’s removal from society \cite{giner2012dehumanization}. Among the common narratives attached to dangerous speech is that of ``accusations in a mirror'', where users will project dangerous behaviour onto target groups \cite{speechdangerous}. 

Dangerous speech, by its nature of dehumanisation and instigation, is overwhelmingly false or exaggerated, making its study relevant to a vast body of research that links disinformation with
political propaganda and polarisation. For example, as \citet{chinmayi2019} points out, incitements to violence in the follow-ups to lynchings have been misconstrued as crises of “fake news” in India - a costly error that transfers culpability from the structural and societal conditions in which violence occurs to the content of the text. In India, concerns about the deterioration of the information environment are well documented, both on traditional news platforms as well as social media. As our research design reflects, the systematic targeting of minorities around contentious events raises important questions about the constructions of in-groups and out-groups in orchestrating and justifying violence. Thus, we aim for computational adaptations of the dangerous speech framework to remain reflexive of the historical and organised
ways in which speech and violence have been linked in the Indian context.

\subsection{Influential Users}
A key attribute of dangerous speech is its propagation by influential users or those whose word is considered more legitimate. Such users with large followings reach broad publics and subsequently generate the largest cascades on Twitter \cite{bakshy2011everyone}. This is backed by the observation that political leaders and influencers across the world \cite{amarasingam2020qanon, castaneda_2019, sippy_2021, martinic_2021}, have played a critical role in the propagation of hate speech, disinformation, and polarisation on social media. There is an ideological skew among influential users, too - in 2021, Twitter discovered that it algorithmically amplifies right-wing elected officials over left-wing elected officials and right-allied news outlets over left-allied news outlets\footnote{https://blog.twitter.com/en\_us/topics/company/2021/rml-politicalcontent}. As Soares, Recuero, and Zago point out, opinion leaders in echo chambers achieve centrality by producing radicalized or biased content that reinforces the political views of the group \cite{soares2019asymmetric}. Accounts that retweet instigators, otherwise known as second speakers, further introduce hateful messaging to newer audiences \cite{speechdangerous}. \citet{dash2021divided} demonstrate that influencers who generated Twitter cascades during the three events explored in this study, were not only consistently polarized across events, but were also rewarded with increased retweets and following for the same, allowing for further dissemination of harmful messaging. This becomes an important discovery in light of the fact that dangerous speech is used by political leaders and influencers towards ends of political polarisation, and allows us to posit it as a tool of strategic propaganda within India’s current political context.

\subsection{India's Political Context}
Our discussion occurs in the backdrop of India’s current socio-political environment, where the right-wing national incumbent Bharatiya Janata Party’s promotion of a hyper-nationalist, Hindu
majoritarian political ideology \cite{chakravartty2015mr, udupa2016archiving, banaji2018vigilante} has led to polarisation and subsequent radicalisation of vote bases. According to the 2011 census, Hindus constitute India’s majority religious demographic at 79.8 \% \cite{india_2011}. Hindutva, a right-wing, populist manifestation of demands from the dominant Hindu community in India  \cite{hansen1999saffron, tillin2017populism}, has othered vulnerable minorities -  namely Muslims, Christians, and Sikhs, who make up 14.2 \%, 2.3 \%, and 1.72 \%  of the country respectively \cite{india_2011} - into categories of ``anti-national'', with clear distinctions drawn between the in-group and the out-group \cite{oommen1986insiders}. Dangerous speech is both a product and driver of this ongoing, technologically mediated political crisis. Indeed, in their work on social media strategies adopted by Indian political parties, \citet{mahapatra2019polarisation} warn that political leaders’ increasing dissemination of polarising content has immense ramifications for democracy.

\section{Methodology} \label{method}
\begin{figure}[!htb]
    \centering
    \includegraphics[width=0.99\textwidth]{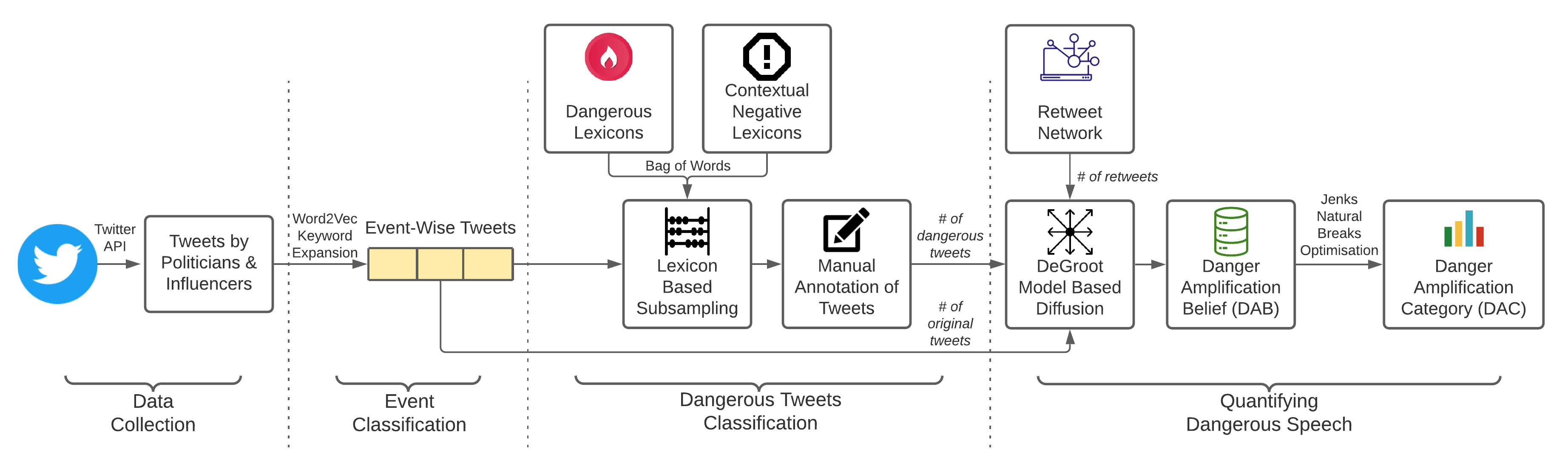}
    \caption{\textbf{Pipeline for Quantifying Dangerous Speech on Twitter}}
    \label{method_pipeline}
\end{figure}

We describe the data collection methodology, quantification of dangerous speech and characterisation of dangerous speech users here. The complete pipeline can be found in Figure \ref{method_pipeline}. The varying attributes of the dangerous speech framework are captured in different stages of the pipeline. The dangerous speech classification mechanism captures the inflammatory nature as well as context of the messaging. The influence of the speaker, which is measured by how much they are endorsed (retweeted) by their audience (retweet network), is incorporated through a belief diffusion process in the quantification of dangerous speech. Moreover, the susceptibility of the audience is measured through derived attributes which indicate how polarized the viewership of the individual's messaging is, which is further symptomatic of their inclination to conflict and violence.

\subsection{Data Collection}

We use the dataset of Indian politicians and influencers used in \citet{dash2021divided}. We briefly describe their data collection methodology here.

Similar to \citet{dash2021divided}, to build the politicians' database, we use a publicly available dataset \cite{PandaNivaDuck} of 36k Indian politicians' Twitter accounts, which include various elected representatives at the state and national level from several parties, as well as party workers like general secretaries, Information Technology (IT) cell heads etc. The \citet{PandaNivaDuck}
dataset is curated by using a Machine Learning classification pipeline called NivaDuck, and manually validated by human annotators. Moreover, the politicians are also manually annotated by the state and party they belong to. We then use the party labels to filter 14,094 BJP politicians and 12,341 INC politicians. 

To build the set of influencers, \citet{dash2021divided} iteratively collect the friends of the 26k BJP and INC politicians. From the resultant list, they remove all the politicians from \citet{PandaNivaDuck}, non-Indian global figures etc., such that they are left with a total of 10k Twitter accounts that are highly followed by Indian politicians. They then manually annotate the filtered accounts to remove the false positives, and categorize each influencer account into one category, whichever they are primarily known by. This results in a total of 6626 influencer accounts. This final set of influencers are divided into 12 categories such as \textit{Academia}, \textit{Journalists} etc. One of the categories that is particularly relevant to our work is the \textit{Platform Celebrity} category which captures a digital phenomenon of accounts whose popularity stems from their online presence and content creation, rather than the offline work of the individuals involved. Refer to \citet{dash2021divided} for more details on the complete list of categories.

Finally, over 43M English tweets are collected\footnote{Using Tweepy - \url{https://www.tweepy.org/}} from these 26k politicians and 6k influencers accounts between June 2019 - March 2021. The tweets are preprocessed in a manner similar to \citet{rashed2020embeddings}, including case-folding and removal of links, emojis, punctuations and non alphanumeric characters.

\subsection{Event Classification of Tweets}
To capture event specific tweets, we use the pipeline in \citet{dash2021divided}, where they use a Word2Vec \cite{mikolov2013efficient} bag of words based technique to classify the collected tweets \cite{vijayaraghavan2016automatic}. Briefly, they first define a set of high precision keywords that are indicative of the event and then train a Word2Vec model based on the tweets that contain at least one of the keywords. The word embeddings are then used to iteratively expand the seed set according to the cosine similarity based criteria defined in \citet{vijayaraghavan2016automatic}. Refer to \citet{dash2021divided} for the specific keywords used to classify the tweets into the different events. For the sake of completion, we include the final distribution of event-specific tweets in Table \ref{topic_stats}.

\begin{table}[!htb]
\centering
\small
\begin{tabular}{{|p{0.15\textwidth}|p{0.1\textwidth}|p{0.1\textwidth}|}}
\hline
\textbf{Topic} & \textbf{Users}  & \textbf{Tweets} \\
\hline
CAA/NRC & 13,744 & 527,702 \\
\hline
COVID-19 & 20,871 & 3,069,851 \\
\hline
Farmers' Protest & 12,224 & 269,742 \\ \hline

\end{tabular}
\vspace{6pt}
\caption{\textbf{Number of Users and Tweets by Event}}
\vspace{-20pt}
\label{topic_stats}
\end{table}

\subsection{Dangerous Speech Classification}
Due to the large volume of tweets for each of the events, we use a lexicon based technique to sample a subset of the event-specific tweets that are likely to be used in an inflammatory context. We further manually annotate the subsample to identify dangerous tweets.
\subsubsection{Lexicon Based Subsampling }
One of the distinguishing aspects of dangerous speech is context, specifically \textit{``social or historical context that has lowered the barriers to violence or made it more acceptable...including previous episodes of violence between the relevant groups''}\cite{speechdangerous}. Therefore, based on India's history of communal violence, we first identify the target groups of dangerous speech across all of the events. This further aids in narrowing down the lexica that could be used in an inflammatory context. For instance, during CAA and COVID-19, the target group for dangerous speech were Muslims, where in the former, the dominant narrative was that the protesters posed a threat to the Hindu community (\textcmttfont{\#CAAIslamistLinkProbe, \#JihadiTerrorists}). In the targeting around COVID-19 the messaging furthered that members of the Muslim community were deliberately trying to spread the virus throughout the nation (\textcmttfont{\#CoronaJihad, \#JihadiVirus}). During the Farmers' Protests rumours alleged that the protesters were trying to revive a secessionist movement in India and therefore posed a threat to the national integrity of the country (\textcmttfont{\#KhalistaniExposed, \#TraitorNotTractor}). Therefore, based on our contextual knowledge of the events, we curate a list of dangerous lexica that are most likely to be used in tweets that fall under the framework of dangerous speech.

Moreover, to filter out tweets where the lexica were used in an innocuous context, for instance condemning the use of some terms like \textcmttfont{Jihadists}, \textcmttfont{Khalistani} etc., we built a set of contextual negative lexica, which when used in the context of the dangerous lexica are most likely counter-speech. For instance, \textcmttfont{dubbed} in
\blockquote{\textit{farmers are being dubbed as khalistani since they launched the protest against farm laws}}
or \textcmttfont{blamed} in
\blockquote{\textit{why is just the tablighijamat being blamed for coronavirus?}} are good candidates for negative lexica. 

Once we curate a set of lexica and negative lexica, some of which are displayed in Table \ref{dangerous_keywords}, we filter out tweets that contain at least one of the dangerous lexica but none of the negative lexica, for further manual annotation. We plan on making the full list of lexica and negative lexica public, for reproducibility.

\begin{table}[!htb]
\centering
\small
\begin{tabular} {|p{0.17\textwidth}|p{0.15\textwidth}|p{0.27\textwidth}|p{0.3\textwidth}|}
    \hline
     \textbf{Event} & \textbf{Target Group} & \textbf{Lexica} & \textbf{Negative Lexica}\\
    \hline
    CAA/NRC & Muslims & parasite,intruder, \textcmttfont{\#jihaditerror} & called, calling, labelling \\
    COVID-19 & Muslims & jihadist,\textcmttfont{\#coronajihad, \#nizamuddinidiots} & islamophobia, blaming, label\\
    Farmers' Protests & Sikhs \& Farmers & khalistani, \textcmttfont{\#khalistaniterrorists, \#traitornottractor} & discredit, dubbed, defame \\
    \hline
\end{tabular}
\vspace{6pt}
\caption{\textbf{List of Target Groups, Lexica and Negative Lexica by Event.}}
\label{dangerous_keywords}
\vspace{-16pt}
\end{table}

\subsubsection{Manual Annotation}
We then manually label each tweet in the subsample of tweets, as dangerous speech or not. Two annotators annotate each tweet, based on the dangerous speech framework in \citet{speechdangerous} as well as their contextual knowledge of the events and the Indian political landscape. We measure inter-annotator reliability using the Cohen's Kappa coefficient. We report scores of 0.92, 0.73 and 0.88 for CAA/NRC, COVID-19 and Farmers' Protests respectively. Due to the highly subjective nature of dangerous speech, we further label a tweet as dangerous only if both annotators agree on the label for the tweet.  The distribution of the number of tweets after subsampling, the number of tweets annotated as dangerous speech and the number of users that engage in dangerous speech are in Table \ref{tweet_dist}.
\begin{table}[!htb]
\centering
\small
\begin{tabular} {|p{0.15\textwidth}|p{0.27\textwidth}|p{0.2\textwidth}|p{0.15\textwidth}|}
\hline
      \textbf{Event} &
      \textbf{\# of Tweets After Subsampling} &
      \textbf{\# of Dangerous Tweets} &
      \textbf{Users} \\
    \hline
    CAA/NRC & 5,323 & 3,949 & 991  \\
    COVID-19 & 1,853 & 1,379 & 573 \\
    Farmers' Protests & 3,047 & 2,265 & 880  \\
    \hline 
\end{tabular}
\vspace{6pt}
\caption{\textbf{Distribution of Dangerous Speech Users and Dangerous Tweets by Event}}
\label{tweet_dist}
\vspace{-20pt}
\end{table}
Note that we define a dangerous user as any user that has tweeted at least one tweet which has been classified as dangerous. The degree of offence, based on attributes like their influence in the network, the frequency of dangerous tweets etc., is calculated below.

\subsection{Quantifying Dangerous Speech}

We use a DeGroot Model \cite{degroot1974reaching,golub2010naive} based diffusion algorithm to quantify the \textit{Danger Amplification Belief} (DAB) score of a user.  \citet{ribeiro2017like} use the diffusion model to select a subset of users to annotate while \citet{mathew2020hate} use it to quantify hate intensity of a user by making use of temporal snapshots of the Gab network. However, as explained previously, any mechanism to quantify dangerous speech needs to take into account nuances like influence of the speaker, which is not vital to the hate speech framework. 

To adapt the DeGroot Learning model to our study, we leverage the induced retweet network and incorporate the influence of the individual by computing how much they are retweeted for their messaging in the network. Furthermore, we take into account the user's proclivity to engage in dangerous speech by initialising the ``belief'' of the model with the user's frequency of publishing dangerous tweets. The term \textit{belief} here in the context of social media platforms can be interpreted as some behaviour that the user adjusts as a consequence of being influenced by their neighbors' behaviour \cite{golub2010naive}. Additionally, we use the volume of tweets of the user as a proxy for their vested interest in the event for pushing forth their agenda \cite{caiani2021online, darius2019hashjacking}.

\vspace{-2pt}
\subsubsection{Danger Amplification Belief Scores}
In the DeGroot model, each user starts with some initial \textit{belief} $p^{(0)}$ which is constantly updated by interacting with their neighbors in accordance with some social network, whose patterns are captured through a $n x n$ nonnegative Transition matrix $T$. 
We extend the model to incorporate the Dangerous Speech parameters.
Let $G = (V,E)$ be the weighted, directed, retweet induced graph of an event, where $V$ is the set of all users who have tweeted about the event and an edge $(u,v) \in E$ if $u$ has retweeted $v$ and $w(u,v) = n_{RT}^{uv}$, where $n_{RT}^{uv}$ is the number of time $u$ has retweeted $v$. Let $A(G)$ be the weighted adjacency matrix, which is defined as: 

\begin{equation}
    A(u,v) = 
    \begin{cases}
    n_{RT}^{uv}, & u \neq v \\
    n_{OG}^u, & u = v 
    \end{cases}
\end{equation}

\noindent where $n_{OG}^u$ is the number of original tweets by $u$ for that event. The transition matrix $T$ is then constructed by taking the transpose of the adjacency matrix \cite{ribeiro2017like,mathew2020hate} (since influence is directed from the retweeted user to the user who retweeted them). Each row of the transition matrix is then normalized such that it sums to 1. Overall, the transition matrix encapsulates the tweeting activity of a user in terms of how much they retweet and publish tweets. The belief vector $p^{(0)}$ is then initialized such that $p_u^{(0)} = n_{DS}^u$, where $n_{DT}^u$ is the number of tweets for user $u$ that are labelled as dangerous speech. While \citet{ribeiro2017like, mathew2020hate} initialize $p^{(0)} \in [0,1]$, we use an alternative initialization of the DeGroot model \cite{golub2010naive}, such that $p^{(0)} \in \mathbb{R}$, in order to factor the frequency of engaging in dangerous speech in the overall score. Finally, new beliefs are updated according to the update rule:

\begin{equation}
    \mathbf{p}^{(t)} = \mathbf{T}\mathbf{p}^{(t-1)}
\end{equation}

Similar to \citet{ribeiro2017like}, we run the diffusion algorithm for $t = 2$ iterations, since the belief values converge to a constant value as $t \to \infty$. The cumulative density function of the computed DAB scores are showed in Figure \ref{cdf}.

\vspace{-2pt}
\subsubsection{Danger Amplification Category}
The DAB score ranges from $[0,1]$, therefore we need a thresholding mechanism to distinguish between the innocuous accounts and the dangerous users. Moreover, it is encouraged to view dangerous speech as a spectrum, rather than a binary attribute -- \textit{``dangerousness can’t be correctly understood as a toggle or light switch, on or off, dangerous or not. Instead it falls on a spectrum...''} \cite{speechdangerous}. Therefore, we consider three danger amplification categories, not dangerous (N), moderately dangerous (M) and very dangerous (V). The moderately dangerous (M) category particularly captures those ambivalent users who are likely to use subtle messaging to instigate their audience.

\begin{figure}[!htb]
\centering
\begin{subfigure}[t]{.5\textwidth}
  \centering
  \includegraphics[width=0.9\linewidth]{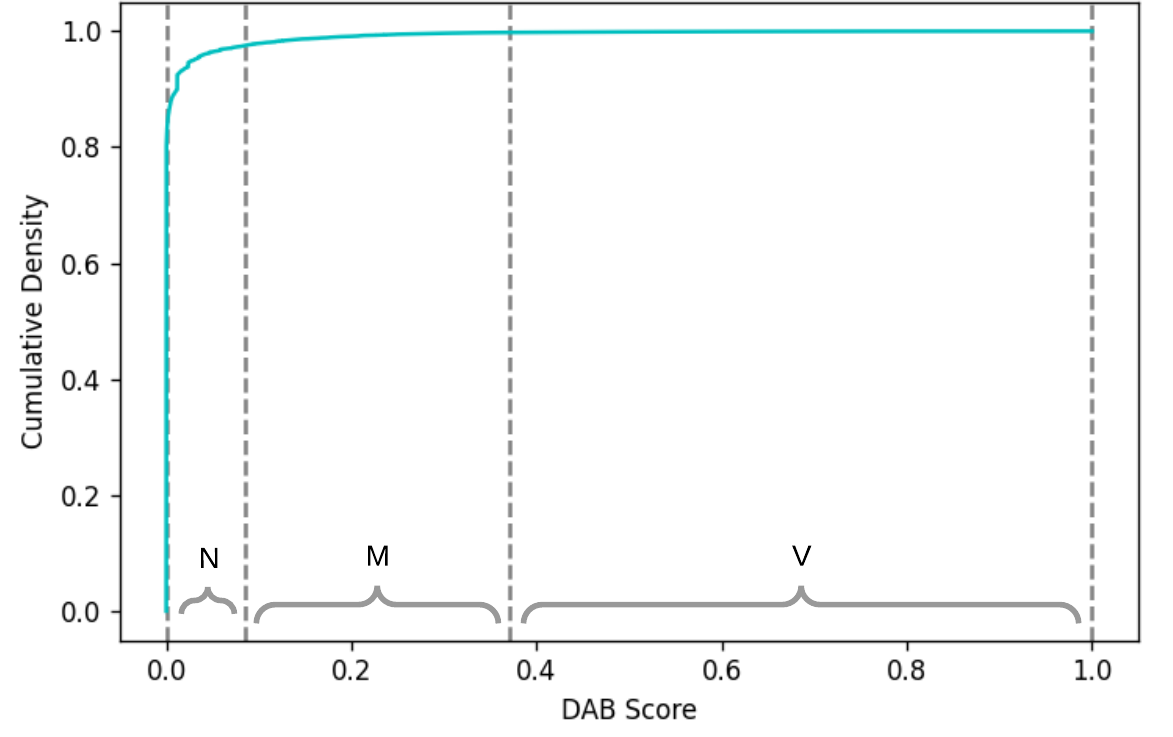}
  \caption{CAA/NRC}
\end{subfigure}%
\begin{subfigure}[t]{.5\textwidth}
  \centering
  \includegraphics[width=0.9\linewidth]{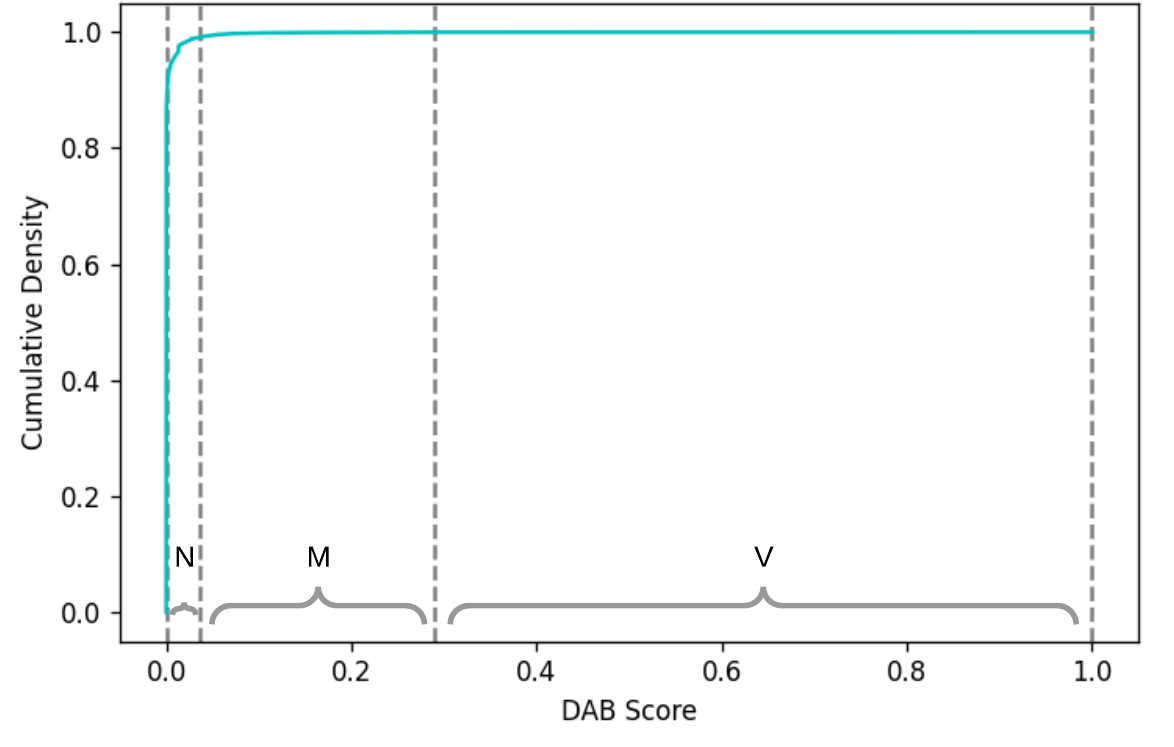}
  \caption{COVID-19}
\end{subfigure}
\begin{subfigure}[t]{.5\textwidth}
  \centering
  \includegraphics[width=0.9\linewidth]{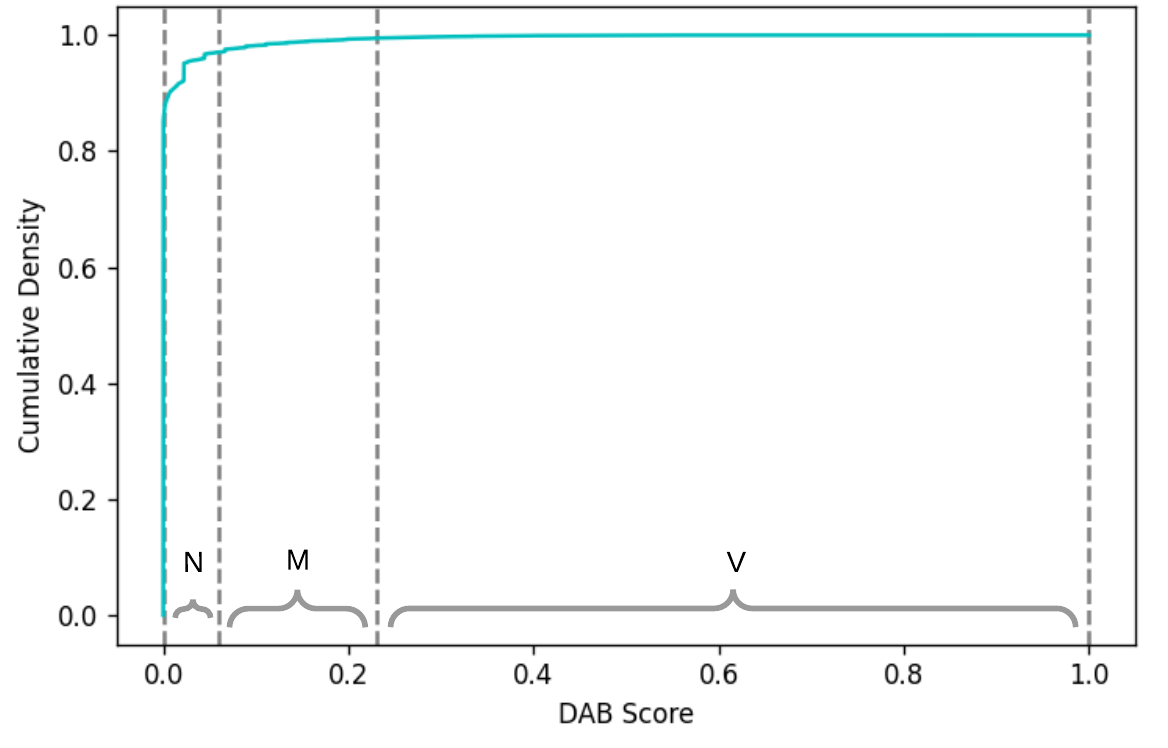}
  \caption{Farmers' Protests}
\end{subfigure}
\caption{\textbf{Danger Amplification Belief (DAB) Distribution \& DAC Thresholds by Event}}
\label{cdf}
\end{figure}

Similar to \citet{mathew2020hate}, we use a clustering algorithm on the DAB scores, to select threshold values between 0 and 1. We use the Jenks Natural Breaks algorithm \cite{jenks1967data} instead of K-Means used by \citet{mathew2020hate}, since K-Means is susceptible to outliers and does not work well with clusters of varying sizes. The Jenks Natural Breaks algorithm optimizes to reduce intra-class variance and increase inter-class variance. The thresholds obtained from the algorithm are displayed in Figure \ref{cdf}. Approximately 4\% of all the users are labelled as dangerous users (M or V) for CAA/NRC, 1\% for COVID-19 and 6\% for Farmers' Protests.

\subsection{Quantifying Polarization}
In order to measure the susceptibility of the individual's audience to inflammatory messaging, we compute polarity scores for the users, which capture different aspects like how polarized is the their retweet network as well as how polarized is their following on Twitter. The retweet network and following act as proxies for the audience of the individual. Moreover, we use polarization as a measure of the susceptibility of the audience to incitement because several studies show that the risk of conflict and violence is strongly correlated with group polarization \cite{harel2020conflict, varshney2008ethnic, weber2013secular}.

\vspace{-4pt}
\subsubsection{Retweet Polarity: }\label{rt_polarity}
Retweets in general correspond to endorsement of the original tweet (excluding quote tweets). The polarizing nature of a tweet can be inferred by the users who endorse it, and whether only individuals from one side of the issue endorse it. Therefore, to characterize the retweets we use the retweet polarity score, which quantifies how likely the user's tweet is to be retweeted by users whose stance on issues is consistently aligned with one of the political parties. The score ranges from [0,1] where scores closer to 0 indicate low polarization and scores closer to 1 indicate high polarization.
We use the methodology in \citet{dash2021divided} to compute the retweet polarity scores. They use Google's Universal Sentence Encoder (USE) \cite{cer2018universal} to encode the tweets and consequently obtain aggregate vector representations of the users based on their tweet embeddings. These alongside retweet graphs help compute the stance and polarity of the users with respect to the political issues. 

\vspace{-4pt}
\subsubsection{Follower Polarity:} Following an account on Twitter indicates a regular viewership of their content. While this does not directly indicate endorsement, a significant group level follower-ship behavior can be a proxy for polarizing content being produced by an account. To quantify this, we obtain the friends of all politicians from our sample, using the Twitter API. Using this we calculate the total number of politicians from BJP and INC following each account in our dataset. We adopt the partisan score presented in Hemphill's  study of partisan hashtags \cite{hemphill2016} and define our score as the Chi-Squared statistic of dependence between number of users that follow an account and their party. Further we set the score to 0 for accounts with p-value >.005 and the log scaled test statistic for other accounts. 

\section{Characterizing Dangerous Speech Users} \label{ds_character}
We analyse how dangerous users differ from the other users in terms of platform attributes such as statuses count, retweet count, followers count etc. and inferred features such as retweet polarity and follower polarity. In order to capture general behavioral trends, we average out the DAB scores for all the users across all the events and recalculate the DAC thresholds using the Jenks Natural Breaks Optimization algorithm.

We use linear regression weights to explore whether the computed DAB scores are indicative of numerical attributes like number of tweets, retweets etc. We also plot the variation of the attributes with the DAC of the users, to infer the trends with respect to the different categories, specifically the ambivalent category of moderately dangerous (M) users. We then analyse the types of dangerous speech users by their influence category and conclude the section by characterizing the position of the users in the social network.

\begin{table}[h]
\centering
\begin{tabular}{|lc}
\hline
\multicolumn{2}{|c|}{\textit{\textbf{y$\sim$dab\_score}}}                     \\ \hline
\textit{log\_statusesCount}       & \multicolumn{1}{c|}{10.02 (p \textless .001)}  \\
\textit{log\_followersCount} & \multicolumn{1}{c|}{12.46 (p \textless .001)} \\
\textit{log\_friendsCount} & \multicolumn{1}{c|}{2.38 (p \textless .001)} \\
\textit{log\_favouritesCount} & \multicolumn{1}{c|}{4.35 (p \textless .001)} \\
\textit{retweetPolarity}     &
\multicolumn{1}{c|}{0.24 (p \textless .01)}   \\
\textit{followerPolarity}    & \multicolumn{1}{c|}{3.71 (p \textless .001)}  \\ \hline
\end{tabular}
\vspace{6pt}
\caption{\textbf{Regression Analysis.} DAB Score (dab\_score : Independent Variable) and Numerical Attributes of Users (y : Dependent Variable)}
\vspace{-8pt}
\label{reg_analysis}
\end{table}

\begin{figure}[h]
\centering
\begin{subfigure}[t]{.25\textwidth}
  \centering
  \includegraphics[width=0.99\linewidth]{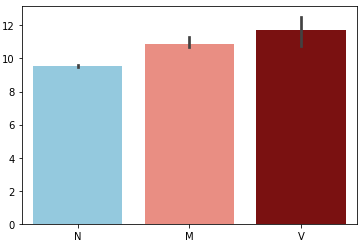}
  \caption{Statuses Count}
  \label{dac_ntweets}
\end{subfigure}%
\begin{subfigure}[t]{.25\textwidth}
  \centering
  \includegraphics[width=0.99\linewidth]{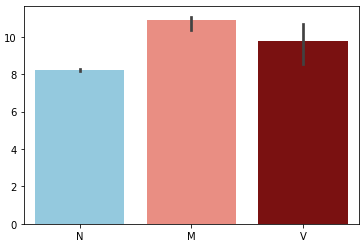}
  \caption{Followers Count}
  \label{dac_fc}
\end{subfigure}%
\begin{subfigure}[t]{.25\textwidth}
  \centering
  \includegraphics[width=0.99\linewidth]{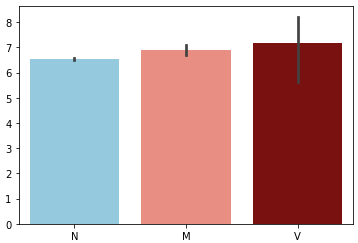}
  \caption{Friends Count}
  \label{dac_fr}
\end{subfigure}%
\begin{subfigure}[t]{.25\textwidth}
  \centering
  \includegraphics[width=0.99\linewidth]{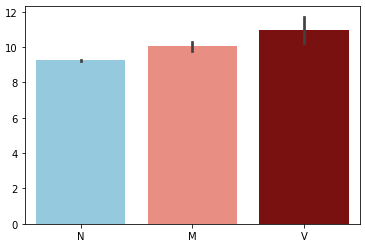}
  \caption{Favourites Count}
  \label{dac_fav}
\end{subfigure}

\begin{subfigure}[t]{.25\textwidth}
  \centering
  \includegraphics[width=0.99\linewidth]{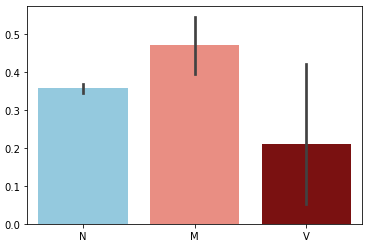}
  \caption{\% of Verified Accounts}
  \label{dac_va}
\end{subfigure}%
\begin{subfigure}[t]{.25\textwidth}
  \centering
  \includegraphics[width=0.99\linewidth]{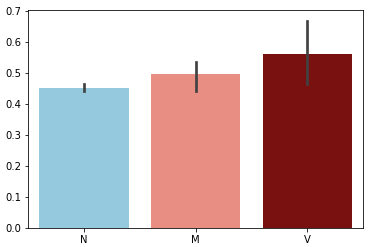}
  \caption{Retweet Polarity}
  \label{dac_rtp}
\end{subfigure}%
\begin{subfigure}[t]{.25\textwidth}
  \centering
  \includegraphics[width=0.99\linewidth]{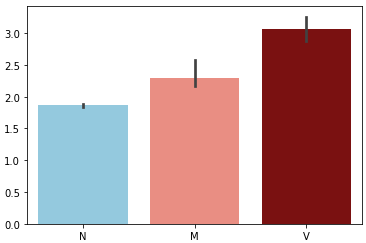}
  \caption{Follower Polarity}
  \label{dac_fp}
\end{subfigure}
\caption{\textbf{Median Values for Platform Attributes Like Statuses Count, Followers Count etc. and Derived Attributes Like Retweet Polarity and Following Polarity by Danger Amplification Category.} Error bar represents the 95\% confidence interval for each distribution.}
\end{figure}

\subsection{Statuses Count}\label{statuses_count}
We start by investigating whether the overall tweeting activity, in terms of number of statuses (tweets), of dangerous speech users differs from other users. We fit a linear regression model with the log-scaled value of the statuses count as the dependent variable and the DAB score as the independent variable. We observe that an increase in the DAB score corresponds to an increase in the number of statuses (slope = 10.02), implying that dangerous users have elevated tweeting activity as compared to other users, similar to hateful users in \citet{ribeiro2017like}. We also visualise how the statuses differ by the different dangerous categories in Figure \ref{dac_ntweets}. We note here as well that there is an increase in the number of statuses as we go from not dangerous (N) users to the very dangerous users (V). We use a one-way ANOVA test to check for statistical differences across the categories and find that with an F-statistic of 101.46 and p-value < 0.001, the tweeting patterns differ significantly across all the categories. Moreover, a post-hoc Honest Significant Difference (HSD) test reveals statistically signifcant differences between each category, i.e. distribution of statuses for not dangerous (N) users is significantly different from (M) and (V) dangerous users, as well as between (M) and (V) dangerous users, implying that (V) dangerous users tweet significantly more than other groups.


\subsection{Followers Count}
In order to analyse the online outreach of the dangerous users in comparison to other users, we compare the log-scaled value of the number of followers of each user with their DAB scores and find a significant positive relation between the two (slope = 12.46). Moreover, we plot the log of the number of followers along with the Danger Amplification Categories (DAC) of the users in Figure \ref{dac_fc}. We find that the (M) category users have higher number of followers as compared to the other two categories. A one-way ANOVA test reveals statistical differences across the categories (F-statistic = 90.22, p-value < 0.001) and a post-hoc HSD test reveals significant differences in following between (N), (M) and (N), (V) users. This finding resonates with the rubrics of dangerous speech that lay emphasis on the influence and power of the user. It can be noted that this is in contrast to the finding in \citet{ribeiro2017like}, where they find that hate speech users are less followed. This further underscores the differences between hateful users and dangerous users.

\subsection{Friends Count}
We compare the log of friends of each user along with their DAB scores, and find that similar to \citet{ribeiro2017like}, there is a positive relation between the two (slope = 2.38), indicating that dangerous users follow more accounts on Twitter as compared to other users. In Figure \ref{dac_fr}, we plot the log of number of friends against the Danger Amplification Category (DAC) of users. We see that the friends count increases marginally as we go from not dangerous (N) to very dangerous (V) category of users. A one-way ANOVA yields an F-statistic of 7.07 and p-value < 0.001, while the HSD post-hoc test indicates statistical differences between (N) and (M) users. This, in conjunction with the followers count analysis, suggests that the dangerous users potentially have a larger network in terms of following and friends. 

\subsection{Favourites Count}\label{favourites_count}
The favourites count attribute of a user on Twitter represents the total number of tweets that the user has liked. This attribute, along with the the statuses count represents platform usage trends. We find a positive relation between the log-scaled value of favourites count of users and their DAB scores (slope = 4.35). We also plot the log of favourites count of users against their Danger Amplification Category (DAC) in Figure \ref{dac_fav}. We note that the favourites count increases from not dangerous (N) users to the very dangerous (V) users, with a one-way ANOVA test indicating statistical differences across the groups (F-statistic = 12.00, p-value < 0.001) and a post-hoc HSD test proving significant differences between (N) users and (M) and (V) users. Therefore, similar to \citet{ribeiro2017like}, we also infer that dangerous users tends to be more active on the platform as compared to other users.

\subsection{Verified Accounts}
As detailed in official Twitter guidelines about verified accounts\footnote{https://help.twitter.com/en/managing-your-account/about-twitter-verified-accounts}, an account must be \textit{authentic}, \textit{noble} and \textit{active}, and must not engage in hateful content, in order to be verified. We plot the percentage of verified accounts for each of the Danger Amplification Categories (DAC) in Figure \ref{dac_va}. In accordance with the Twitter guidelines, very dangerous (V) users have the least percentage of verified accounts. However, it is interesting to note that the moderately dangerous users (M) have the highest percentage of verified accounts as compared to the other two categories. Moreover, a one-way ANOVA followed by a post-hoc HSD test reveals significant differences between (M) and (N), (V) categories (F-statistic = 5.80, p-value < 0.05). This may indicate that while the verified accounts may not directly engage in hate speech, in fear of losing their verified status or getting suspended from Twitter, they are not penalised for publishing dangerous speech, where their targeting is framed in a more subtle manner and open to interpretation, thereby escaping liability for the repercussions of their speech.


\subsection{Retweet Polarity}
We also analyse whether the retweet networks of the dangerous users are more polarized as compared to the other users in the sample. As explained previously, retweets in general correspond to endorsements, with the retweet network offering some insights into the audience of the messaging, and the polarity of the retweet network representing the susceptibility of the audience to inflammatory content.
A positive regression weight between the retweet polarity of the users and their DAB scores (slope = 0.24) offers some insight. To supplement this, we also plot how the retweet polarity of users vary with their DAC in Figure \ref{dac_rtp}. We see that the retweet polarity increases as we go from not dangerous (N) users to moderately dangerous (M) users, peaking at the very (V) dangerous users. This lends to the intuitive understanding that the group of users who endorse the dangerous users are more polarized as compared to those who endorse other users. While a one-way ANOVA does not result in statistically significant differences across the categories, the regression results warrant further investigation into these pattern.

\subsection{Follower Polarity}
Another proxy for the audience of a user's messaging is their following. A significant group level follower polarization offers further insights into the susceptibility of the audience to the user's tweets. A significant positive slope (3.71) from the regression model of the follower polarity dependent on the DAB score, along with the distribution of the the follower polarity of a user against their DAC as shown in Figure \ref{dac_fp} shows that follower-ship polarity steadily increases from the not dangerous (N) to the very dangerous (V) users. Moreover, a one-way ANOVA (F-statistic = 14.01, p-value < 0.001) in conjunction with a post-hoc HSD test reveals significant differences between (N) and (M) users. These trends, along with the retweet polarity patterns suggest that the regular audience of the dangerous users are polarized and hence susceptible to their inflammatory messaging. 

\subsection{Influence Category}\label{influence_cat}
\begin{figure}[!htb]
\begin{subfigure}[t]{.35\textwidth}
  \centering
  \includegraphics[width=0.99\linewidth]{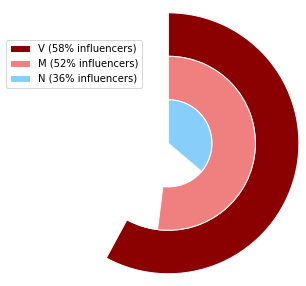}
  \caption{Influencers Percentage by DAC}
  \label{inf_percentage}
\end{subfigure}%
\begin{subfigure}[t]{.65\textwidth}
  \centering
  \includegraphics[width=0.9\linewidth]{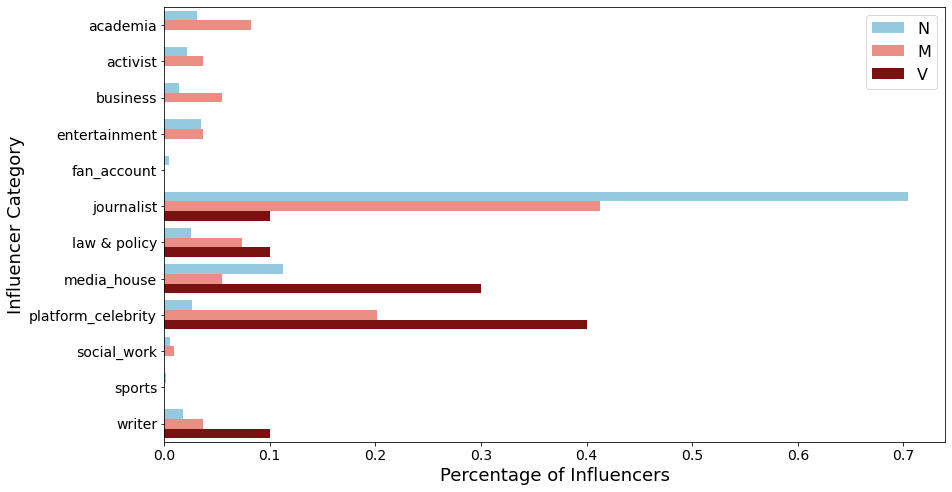}
  \caption{Influencers Category by DAC}
  \label{inf_cat}
\end{subfigure}%
\caption{\textbf{Percentage and Category of Influencers by Danger Amplification Category (DAC)}}
\end{figure}

Here, we attempt to compare the influence category of the users with their Danger Amplification Categories (DAC). In the case of influencers, their influence category could be their offline occupation like journalist, sportsperson etc., or solely their online presence such as the case of platform celebrities. For politicians, identifying their position in the party is slightly trickier. To overcome this, we use their status descriptions as proxies.

In Figure \ref{inf_percentage}, we plot the percentage of users who are influencers in each DAC. We see that the percentage of influencers is highest for the very dangerous (V) category, followed by the moderately dangerous (M) users, and is the lowest for the not dangerous (N) category. This indicates that majority of the dangerous users are influencers.
Furthermore, in Figure \ref{inf_cat} we see that majority of the influencers belong to categories related to mass media including journalists, media houses, platform celebrities and writers. We see that media houses have become an important and institutional component of dangerous speech, and that platform celebrities have an outsize effect on dangerous speech, which suggests that many of these may in fact be politically motivated and managing online profiles with a mandate to stir up passions. Other categories, including academia, activists, business, entertainment, fan accounts, social workers and sportsperson, have very less or no representation in dangerous speech. 

\begin{figure}
    \centering
    \includegraphics[width = 0.6\linewidth]{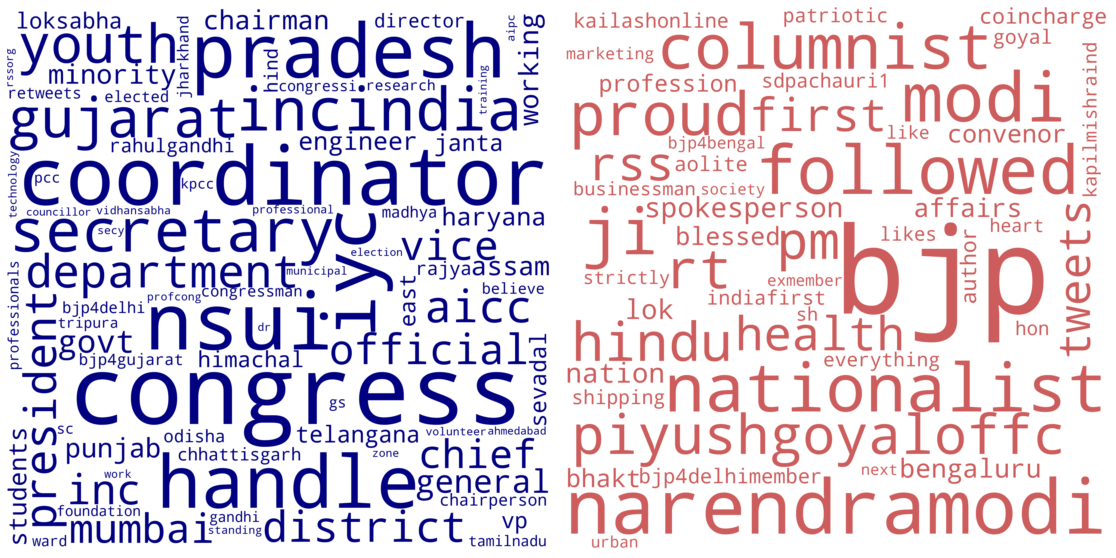}
    \caption{\textbf{Profile Descriptions of Not Dangerous (Left) and Dangerous (Right) Politicians}}
    \label{pol_desc}
\end{figure}

To understand the occupations and categories of politicians, we compare the WordClouds of their status descriptions on Twitter in Figure \ref{pol_desc}. The party leanings are very prominent in the figure. We notice the presence of words relating to the opposition party such as ``congress'', ``iyc'' and ``incindia'' on the ``not dangerous side''. On the other hand, we see words relating to the ruling party like ``bjp'', ``narendramodi'' and ``modi'' on the Dangerous side. It is also interesting to note the presence of patriotic terms like ``nationalist'', ``hindu'', ``patriotic'', ``bhakt'', ``indiafirst'' and ``nation'' on the dangerous side. What we find essentially is a remarkable collection of terms that directly associates accounts owing allegiance to the ruling elite, or with self-assigned terms of nationalism, with engaging dangerous speech on social media. 

In terms of occupations, the dangerous side has lesser words as compared to the not dangerous side. Some of the occupations noted on the dangerous side are ``columnist'', ``spokesperson'' and ``businessman'', while on the other side we notice more ministerial occupations like ``coordinator'', ``secretary'', ``president'' etc. It is also important to note that one of the most common terms of profile text among those indulging in dangerous speech is ``followed by'' typically referring to a major politician. This suggests that the account engaging in dangerous speech seeks legitimacy by referencing a connection with a politician.

\subsection{Network Centrality}
\begin{figure}[!htb]
\centering
\begin{subfigure}[t]{.33\textwidth}
  \centering
  \includegraphics[width=0.99\linewidth]{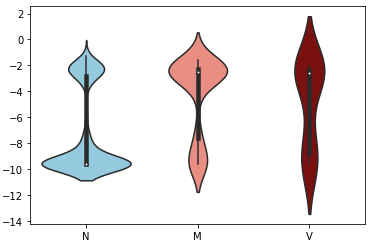}
  \caption{Closeness Centrality}
  \label{closeness}
\end{subfigure}%
\begin{subfigure}[t]{.33\textwidth}
  \centering
  \includegraphics[width=0.99\linewidth]{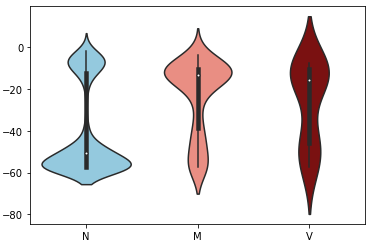}
  \caption{Eigenvector Centrality}
  \label{eigen_centrality}
\end{subfigure}%
\begin{subfigure}[t]{.33\textwidth}
  \centering
  \includegraphics[width=0.99\linewidth]{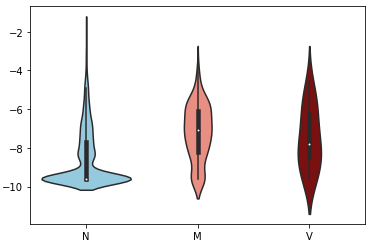}
  \caption{Indegree Centrality}
  \label{fin_degree}
\end{subfigure}
    \caption{\textbf{Distribution Values for Closeness Centrality, Eigenvector Centrality and Indegree Centrality in the Retweet Induced Graph by Danger Amplification Categories.} Values are log scaled and white dot represents the median. Dangerous Users have significantly higher centrality measures as compared to other users.}
    \label{network_centrality_measures}
\end{figure}

To understand the location and importance of dangerous users amongst our sample, we use the retweet induced graph described in Section \ref{method} to calculate three centrality scores across the events and present distributions in Figure \ref{network_centrality_measures}.  We see that (M) accounts generally have high centrality scores, including a significantly higher eigenvector score in comparison to the other categories. A high eigenvector centrality indicates a high probability of being retweeted by other highly central accounts in the sample. This along with a high in degree measure can be explained by the high followers count and verified accounts of the (M) group as seen in Figures \ref{dac_fc} and \ref{dac_va}. The (M) and (V) users also have high closeness centrality. A high closeness centrality corresponds to nodes that have the shortest paths to other nodes on an average, suggesting an ability to share information very quickly and more directly. This suggests that the dangerous users are not only highly retweeted but are also well situated in the network to disseminate their messaging quickly.  

\begin{figure}[!htb]
\centering
\begin{subfigure}[t]{.33\textwidth}
  \centering
  \includegraphics[width=0.99\linewidth]{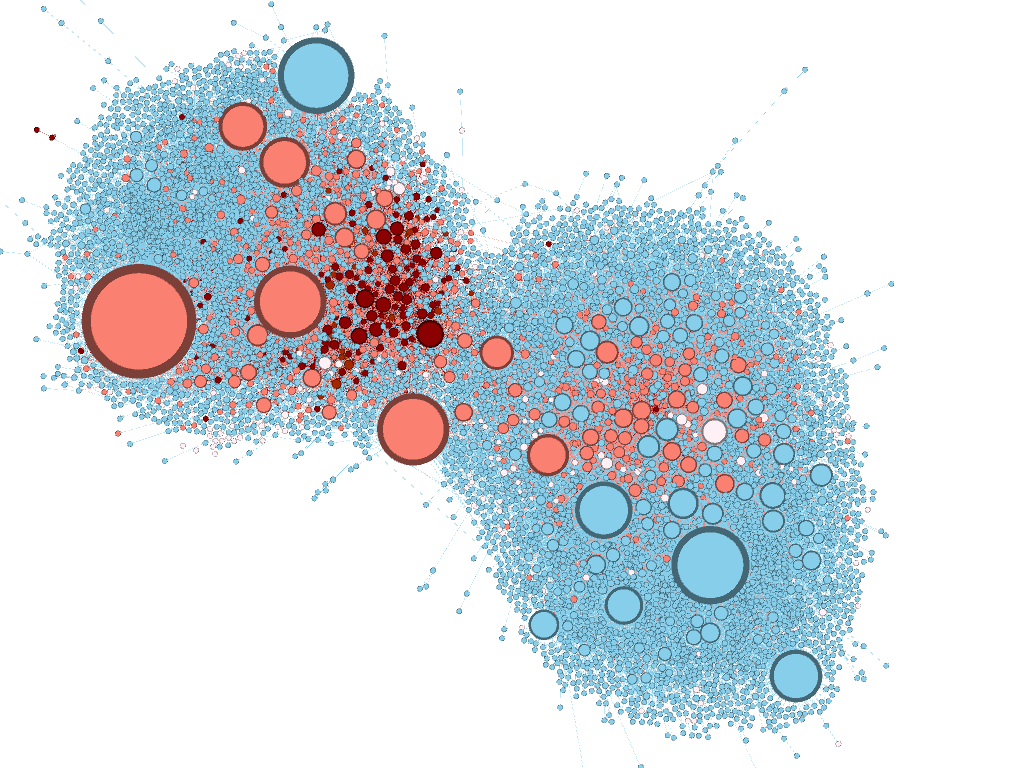}
  \caption{CAA/NRC}
  \label{caa_rt_network}
\end{subfigure}%
\begin{subfigure}[t]{.33\textwidth}
  \centering
  \includegraphics[width=0.99\linewidth]{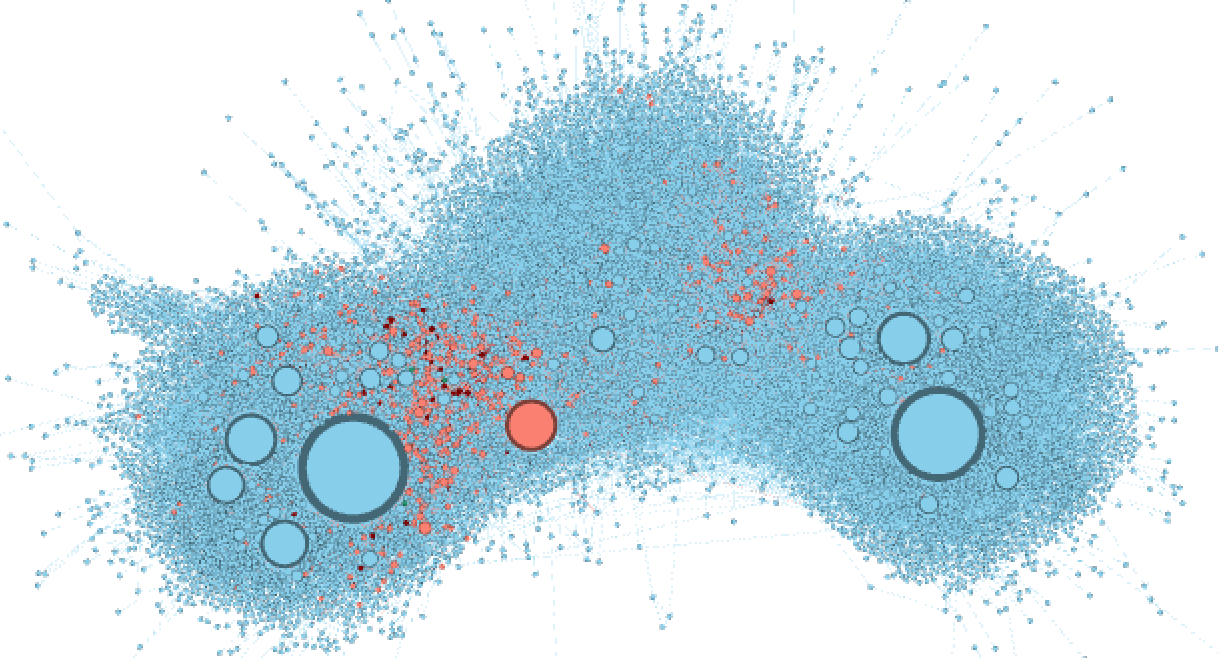}
  \caption{COVID-19}
  \label{covid_rt_network}
\end{subfigure}%
\begin{subfigure}[t]{.33\textwidth}
  \centering
  \includegraphics[width=0.99\linewidth]{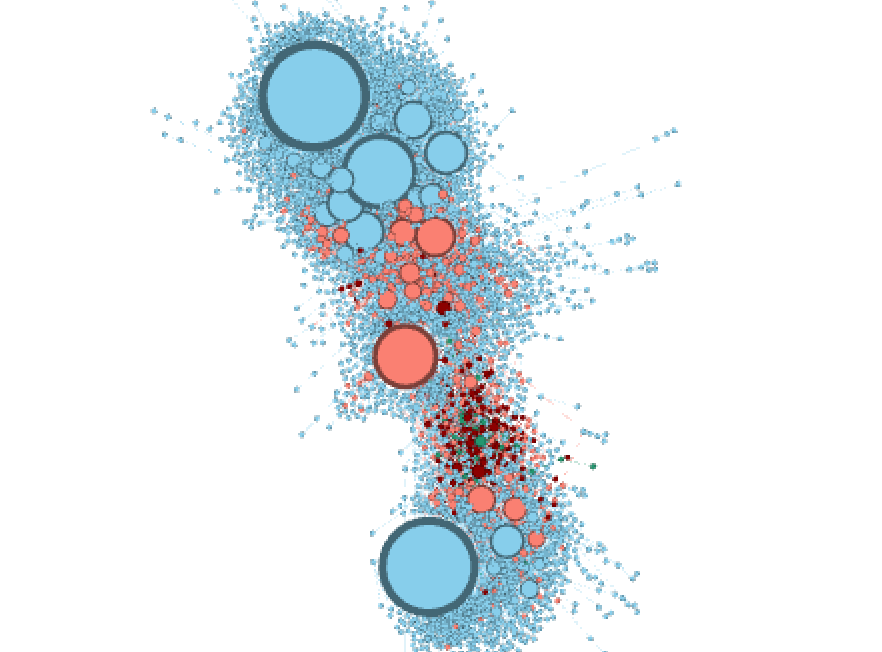}
  \caption{Farmer's Protests}
  \label{farmers_rt_network}
\end{subfigure}
\caption{\textbf{Retweet Networks of the Polarizing Events Color Coded by Danger Amplification Categories.} Sky blue, salmon and dark red colors represent not dangerous (N), moderately dangerous (M) and very dangerous (V) categories respectively.}
\label{rt_network}
\end{figure}

We further visualise the retweet induced graph using the Force Atlas 2 algorithm on Gephi in Figure \ref{rt_network}. The gravity based algorithm is able to clarify the sample into two distinct clusters for each event with a cluster connecting both of them. Color coding these nodes based on their DAC categories, shows that most (V) users occupy the more central positions in the graphs as was suggested by their high closeness centrality albeit always leaning to one side of the cluster. (M) users not only take up the most central positions on the graph but also make up some of the largest nodes in the cluster indicating high indegree as found earlier. 

\begin{figure*}[!htb]
\centering
\begin{subfigure}{.5\textwidth}
  \centering
 \includegraphics[width=0.9\linewidth]{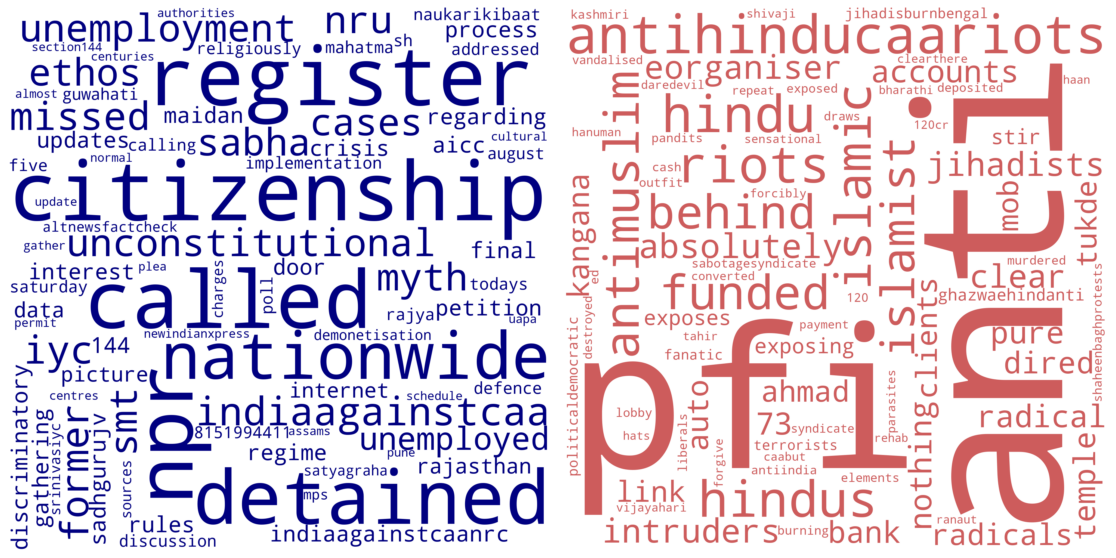}
  \caption{CAA/NRC}
\end{subfigure}%
\begin{subfigure}{.5\textwidth}
  \centering
  \includegraphics[width=0.9\linewidth]{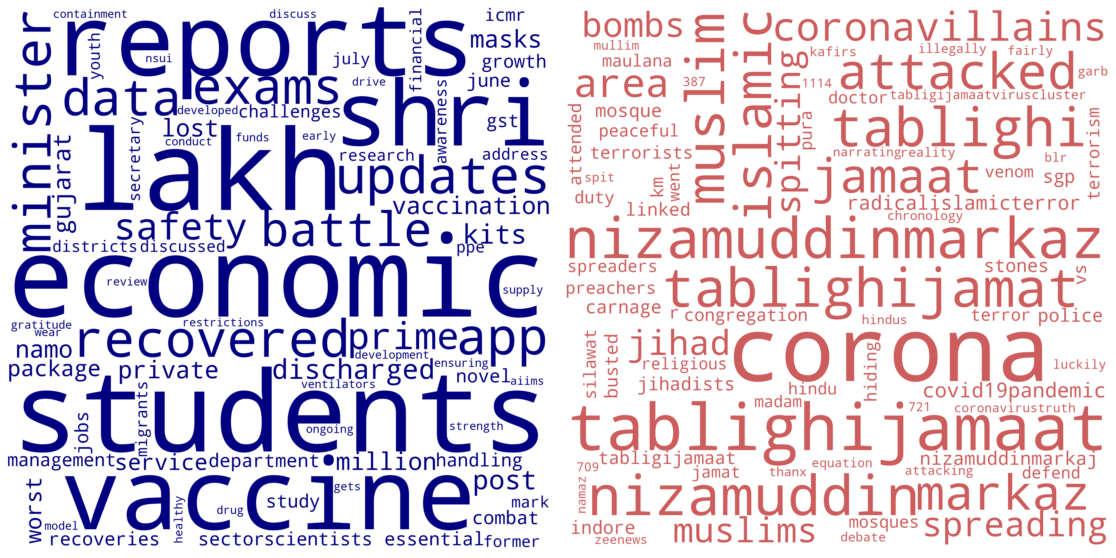}
  \caption{COVID-19}
\end{subfigure}
\begin{subfigure}{\textwidth}
  \centering
 \includegraphics[width=0.5\linewidth]{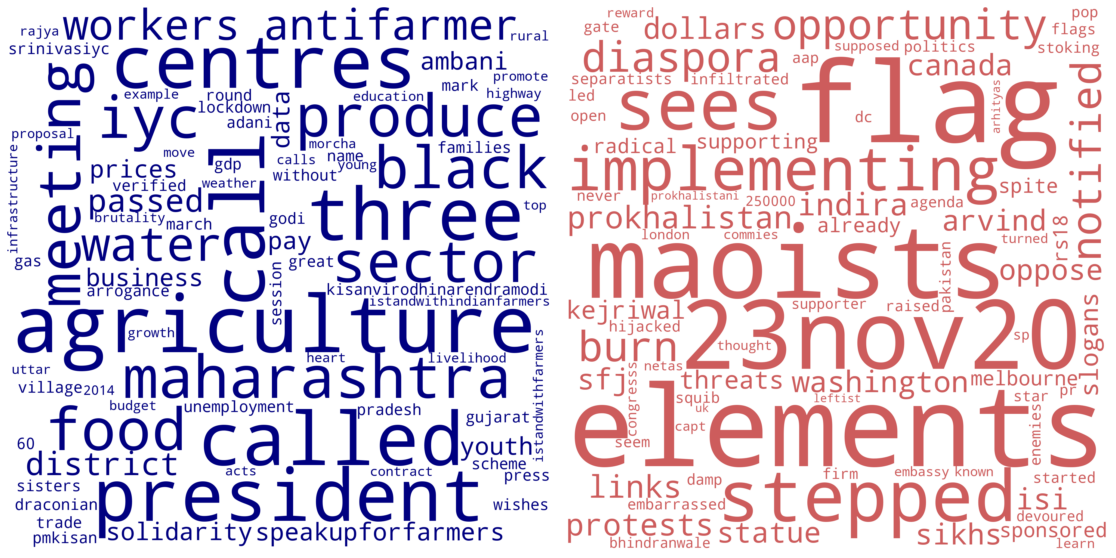}
  \caption{Farmers' Protests}
\end{subfigure}
\caption{\textbf{WordClouds of Not Dangerous Tweets (Left) and Dangerous Tweets (Right) by Event}}
\label{wordcloud}
\vspace{-11pt}
\end{figure*}

\section{Contextualising dangerous speech}\label{ds_content}
In this section, we discuss and contextualise (a) characteristics of dangerous users (moderately and very) via a ground-truth analysis, (b) dominant dangerous speech narratives that emerge within unique datasets of tweets across each of the three events, along with frequently observed keywords and hashtags, and (c) overarching patterns that inform our understanding of both content and political strategy surrounding dangerous speech in the Indian context. Observed together, this qualitative analysis provides key insights into the real-life impact of dangerous speech, as well as how India’s ongoing political circumstance informs dangerous speech on Twitter.

\subsection{Dangerous Users Characteristics}
We examine the Twitter accounts of all 16 very dangerous users as well as the top 20 moderately dangerous users (excluding 2 accounts suspended at the time of authorship). As noted in the word clouds in Figure \ref{wordcloud}, all identified users clearly and uniformly signal affiliation with or support for one or more of three themes -- (a) Indian nationalism, (b) religious majoritarianism, or (c) the BJP and/or the RSS. Collective identity formation towards a common cause plays a key role in the processes of groupism, polarisation, and consequent radicalisation against minorities.

At least ten users, including one Union Minister, one National Spokesperson, and multiple IT, social media, and district level workers, are professionally affiliated with the BJP or RSS.  We note, however, that only three “very dangerous” users describe some explicit role with the ruling party. The rest instead fall under the category of the platform celebrity, whose propagation of dangerous speech is a non conventional, powerful means of influence that allows them to guide viewers towards partisan narratives. Such celebrification of everyday users introduces the idea that the larger political agenda is not merely top-down. Instead, civilian opinion leaders support this viewpoint, giving it an organic legitimacy. These users also proclaim support or “bhakti” (devotion) for the ruling party, or boast a following by Prime Minister Narendra Modi or other influential political leaders -- a signal, as noted in Section \ref{ds_character}, of endorsement for the user or their tweet content. Across bios, alignment with the Indian nationalist cause is signalled through phrases and hashtags such as \textcmttfont{``Bharat Mata Ki Jai''} (hail mother India), \textcmttfont{``Akhand Bharat''} (the promise of a greater India that would subsume most countries in south Asia), \textcmttfont{nationalist}, \textcmttfont{hyper-nationalist}, \textcmttfont{\#ProudIndian}, expertise in national interest and security, a singular Indian flag emojis, or a red flag emoji (a popular signal of support toward Hindu nationalism). We also note a conflation of Hindu identity with Indian identity, with one bio reading, \textit{``Hindu Hai Toh Hindustan Hai''} (if Hinduism exists, India exists), and another stating, \textit{``No compromise on Hindutva''} (no compromise on Hindu nationalism). These findings become particularly relevant in light of the dominant dangerous speech narratives discussed below.

Section \ref{influence_cat} notes that a significant portion of dangerous speech influencers belong to mass-media or journalism related professions. Three influential, hyper-partisan right-wing media outlets, OpIndia, Swarajya  Mag, and eOrganiser, along with the editor of eOrganiser, find a place in the very dangerous user dataset. A similar outlet, OneIndia News, along with one OneIndia editor and one OpIndia editor, are noted in the moderately dangerous category. No user in the very dangerous category apart from the 3 media outlets is verified. While this fact is in line with Twitter’s policy of providing blue ticks to media outlets including digital news publishers\footnote{https://help.twitter.com/en/managing-your-account/about-twitter-verified-accounts}, it also counters Twitter’s claim that the micro-blogging site does not verify individuals or groups that are associated with hateful content. On the other hand, 8 users in the moderately dangerous category have verified profiles. As noted in our quantitative findings, this may be because moderately dangerous users are more likely to rely on innuendo, endorsements, and the garb of social commentary to propagate dangerous speech. For example, one tweet from the moderately dangerous dataset notes, \textit{``there is nothing anti muslim about caa but the anti caa protests were antihindu, tejasvisurya in parliament as he recalls some of the anti hindu slogans raised during the protests''}. The tweet does not include a direct attack against a target community, but when read in the context of the user profile, we note that it propagates a key dangerous speech narrative surrounding the false attribution of anti-Hindu intent to anti-CAA protesters through an endorsement of BJP Member of Parliament Tejasvi Surya. However, we also note that users from both categories routinely publish tweets including keywords such as \textcmttfont{jihadi}, \textcmttfont{radical islamic}, \textcmttfont{khalistani}, as well as hashtags used to popularise dominant dangerous speech narratives.

While users in both very dangerous and moderately dangerous categories claim to be primarily employed via politics, business, entrepreneurship, law, and social activism, their Twitter feeds consist overwhelmingly of original tweets, retweets, and favorited tweets surrounding hyper-partisan political content and dangerous speech towards minorities.  We note both endorsement of content published by official government sources and a distinct lack of representation of non-Hindu identities among retweets. Moreover, all users within this dataset routinely retweet accounts that publish similar tweets, and as discussed in Sections \ref{statuses_count} and \ref{favourites_count}, continue to be considerably active on Twitter. In fact, the volume of tweets posted by users within our dataset verges on bizarre, with accounts such as @56perumal having sent out millions of tweets, and others such as @KapilKrSingh, with just 308 followers on a 5-year old account, having sent out 26,000. Though the authors refrain from claiming proof of coordination, we do note a striking similarity of narratives, strongly suggesting some form of deliberate activity, derailing, or manipulative behaviour \cite{neyazi2020digital}. This also puts to question the failings of Twitter's moderation policies, because repeat dangerous speech users with hundreds of thousands of followers continue to enjoy a large platform. These observations, along with the subsequent echo-chambering of both political ideology and dangerous speech narratives within this information ecosystem, mirror high retweet polarity scores found among very dangerous and moderately dangerous users.

\subsection{Analysis of Dangerous Tweets: Dominant Narratives and Keywords}
We consider both first speakers in the form of original posters, and second speakers in the form of users who retweet and deliver the message to newer audiences. The highest number of tweets marked dangerous are for CAA (n=3949), followed by farmers protests (n= 2265) and COVID-19 (n= 1379). The least number of messages for COVID-19 is due to the specificity of dangerous speech around the Tablighi Jamaat subset, while the prominence of dangerous speech during anti-CAA protests is attributable to the central role that Muslim othering has played in recent manifestations of Indian unity \cite{kumar2013constructing}. In order to reflect the linkages between right-wing nationalism and anti-minority sentiments, we provide an overview of popular narratives found within individual tweets of CAA, Talighi Jamaat, and Farmers' Protests, identified as dangerous speech.


\subsubsection{Anti-Citizenship Amendment Act Protests}
Communalism emerged as central to dangerous speech surrounding the anti-CAA protests. On one hand, tweets accompanied by keywords such as  \textcmttfont{jihadists},  \textcmttfont{islamists}, and  \textcmttfont{radicals}, noted across dangerous speech narratives, heavily attributed Islamic extremism to protestors demanding the revocation of the Citizenship Amendment Act. On the other hand, dangerous speech users posited the larger anti-CAA movement as explicitly ``anti-Hindu'', with keyword  \textcmttfont{anti-Hindu CAA} gaining popularity through the duration of protests. This strategic juxtapositioning of ``anti-national'', Islamic extremism against ``nationalist'', majoritarian Hinduism contributed to the stark religious polarisation seen in this time period, setting the stage for the othering of non-Hindu minorities, and specifically Muslims, across consequent narratives. 

Given that Muslim citizenship was a central concern of the CAA, dangerous speech users used Twitter to further fuel demographic panic among non-Muslim citizens. First, the conflation of ``Muslim'' and ``jihadi'' by users in our dataset painted members of the Muslim community as a threat to the security of the Indian State. Then, this vilification was supplanted with concerns of a rise in the presence of Muslims, and consequently Islamic extremism, in India. We note tweets surrounding this alleged infiltration of Muslims alongside particularly dehumanising keywords such as  \textcmttfont{intruders} and  \textcmttfont{parasites}. The hashtag  \textcmttfont{\#IndiaSupportsCAA}, popularised against the hashtag \textcmttfont{\#IndiaRejectsCAA}, frequently accompanied such tweets. 

A third, popular narrative within our dataset alleged the funding of protestors, lawyers, activists, and broader protest-site logistics by radical Islamic, ``anti-national elements'' -- an idea that conspicuously sought to delegitimize the anti-CAA movement and its many moving parts. The organisation Popular Front of India, the opposition party INC, and the \textcmttfont{tukde tukde gang} (break India gang) -- a pejorative catchphrase used by BJP leaders and supporters in reference to allegedly seditious dissenters \cite{tripathi_2021}, were among such sources. The use of  \textcmttfont{PFIkajihad}, \textcmttfont{radical islamic PFI}, and \textcmttfont{tukdefundedcaastir} shed light on the attribution of extremism in this context. Also pivotal to dangerous speech narratives was Shaheen Bagh, a peaceful protest sit-in led by Muslim women. Hashtags such as \textcmttfont{\#shaheenbaghsham}, \textcmttfont{\#shaheenbaghtruth}, and \textcmttfont{\#shaheenbaghscam} embellished claims that protestors were being funded by the aforementioned sources, misrepresenting protest turnout, or indulging in Islamic extremist conspiracy against the nation. One such tweet reads, \textit{``shaheen bagh is blot on indian democratic history, anti india slogans are being raised...protesting for pak muslims its very shameful disgusting delhi is suffering on daily basis still no action on these islamic jihadis''}. 
Lastly, the use of derogatory ``nicknames'' such as \textcmttfont{urban naxals}, \textcmttfont{sickular} (secular), and \textcmttfont{commie} (communist), coined by right-allied influencers during previous protests and movements, gained popularity - a phenomenon that we discuss later in the section.

\subsubsection{The Tablighi Jamaat Congregation}
The Tablighi Jamaat incident occurred, as dangerous speech surrounding the anti-CAA protests was beginning to simmer. Given this context, dangerous speech users reverted to allegations of jihad related extremism and Islamic radicalisation against Tablighi Jamaat, the Sunni missionary group, and Nizammudin Markaz, the mosque where they congregated. Tweets by users within our dataset broadly alleged that members of the Muslim sect, and extremists within the Muslim community at large, were strategically leveraging the COVID-19 pandemic in order to destabilize the Indian State. The attributed intent here was, as noted in the prior event, to abet the infiltration of Islamic values into the country. An expansive set of keywords and hashtags, including \textcmttfont{\#coronajihad}, \textcmttfont{radical islamic terror}, \textcmttfont{\#jihadivirus}, \textcmttfont{\#tablighivirus}, \textcmttfont{\#coronavillains}, and \textcmttfont{\#indiafightsjihad} points to the blunt, targeted nature of this narrative. 

The ongoing religious gathering in the midst of a call for a nationwide lockdown led to tweets attributing violent intent, alongside keywords such as \textcmttfont{bomb}, \textcmttfont{stones} \textcmttfont{attacking}, \textcmttfont{carnage}, \textcmttfont{weapon} and \textcmttfont{terrorist}, aided the nationwide scapegoating of the Muslim community. Such claims were decidedly direct, with one tweet stating, \textit{“I always believed islamic terrorists will use covid19 as a weapon to try and wipeout as many nonmuslims as possible around the globe this jihadivirus is urging his fellow muslims to spread coronavirus what a disgrace to humanity looks like a new form of jihad coronajihad”}. Further, in light of rising anxieties surrounding physical health during the COVID-19 pandemic, dangerous speech users alleged that the Muslim sect was actively engaged in ``bio jihad'', i.e. biological warfare, with the aim of terrorising citizens and overhauling India on Islamic lines, which has been a long-standing trope of the radical right in India \cite{chaturvedi2002process}. One such example is \textcmttfont{coronabomb}, a term used to popularise the metaphor of suicide bombers, in that Muslims diagnosed with COVID-19 intended to use their bodies as weapons against Indian citizens. Their alleged strategies of choice were \textcmttfont{spitting jihad} and \textcmttfont{defecation}, a dehumanising narrative insinuating a lack of hygiene within the Muslim sect, in contrast to the ``purity'' of the nationalist majority. Another example is one of many widely forwarded messages on Twitter and WhatsApp \cite{akbar2021misinformation} claiming that Muslim street vendors deliberately spat on produce in order to infect Indians with COVID-19. When observed in comparison to the CAA, a distinction in the use of ``jihad'' here is its evocation of capture, which is proposed as deeper and more damaging than simply working against the interests of Hindus. 

Amidst narratives of the sect’s strategic non-cooperation with police, state, and national guidelines to stop the spread of COVID-19, as well as those of linkages between the spread of COVID-19 and Islamic extremism, dangerous speech surrounding the Tabhligi Jamaat incident ultimately served to invoke a rise in anti-Muslim sentiments and consequent political polarisation in India.

\subsubsection{Farmers' Protests}
Dangerous speech propagated by users during the anti-farm bill farmers’ protests diverged from that seen in the two earlier events, since Sikhs and farmers, not Muslims, constituted the target group. Regardless, it remained consistent with allegations of anti-national intent. 

Dangerous speech users labelled protesting Sikhs as \textcmttfont{khalistanis} through the use of keywords including \textcmttfont{pro khalistan}, \textcmttfont{radical}, and \textcmttfont{separatists}. As an example, on India’s Republic Day on January 26 2021, dangerous speech users stoked fears that Sikh farmers had stormed the Red Fort, a monument situated in the country’s capital, and hoisted a Khalistani flag on it. However, the flag was that of Nishan Sahib, representing the Sikh religious community, and the Red Fort was the chosen site of a peaceful sit-in protest\footnote{https://thewire.in/communalism/fact-check-red-fort-flag-tricolour-khalistan-nishan-sahib}. Though the farmers’ protests are decidedly detached from the Khalistani legacy of secessionism and counterinsurgency among Sikhs in the Indian state of Punjab, this narrative created and consolidated fears of the minority community’s incompatibility with a broader nationalist narrative. This targeting of Sikhs is particularly alarming as it brings another religious community to the centre-stage of nationalist ire in a manner, which in contemporary times is usually reserved for Muslims -- a signal that the boundaries of the Indian nation-state are being redrawn to include only the majoritarian, Hindu nation. 

On the other hand, as an occupational group, farmers have been venerated in popular culture and public discourse with slogans such as, \textit{“Hail the soldier, Hail the farmer”}. A strategic skirting of speech around demonisation of the ``farmer'' identity corresponds with the backlash such claims may invite. Instead, another central narrative that emerges is that of ``anti- national elements'' posing as farmers in order to promote anti-establishment protests. This allegation is supplemented by dominant narratives surrounding farmers’ lack of understanding of farm bills, owing to misinformation propagated by "anti-establishment" media outlets, celebrities, khalistani extremists, and myriad international forces. Tweets with accompanying keywords such as \textcmttfont{\#FarmersProtestHijacked}, \textcmttfont{\#ToolkitExposed}, and \textcmttfont{\#TraitorsNotTractors} alluded to the idea that larger, external forces -- from the opposition INC, to the Sikh diaspora, to Rihanna and Greta Thunberg -- sought to destabilise the Indian state. For example, one tweet reads, \textit{“exposed  pop star rihanna was paid over rs18 crores in dollars by pr firm with khalistani links to tweet in support of farmersprotest  18 crore for a single tweet  indiarejectspropaganda”}. Of particular frequency here is the link drawn by dangerous speech users between protesting farmers and Maoism. This narrative seeks to conflate the largely peaceful farmers’ protests and sit-ins with the anti- Indian state, militant approach of Indian Maoists -- a fact noted in light of the designation of the Communist Party of India (Maoist) as a terrorist organisation. 
 
Overall, we observe that dangerous speech during the anti-CAA protests foremost paints them to be form of religious fanaticism, in opposition to Hindus. Therefore, the speech carries incitements to communal violence. The territorial integrity of India is also cited to be under threat, but only second to the threat Muslims pose to Hindus. During COVID-19, the targeting becomes one of biowarfare, where Muslims are seen as human ``bombs'', destroying those around them, in imagery that is akin to suicide bombings. Farmers protests foremost raise the accusation of a secessionist movement, led by those who wish to see India broken up. Indirectly, it targets Sikh protesters and questions their loyalties, while avoiding attacks to their occupational identity.

\subsection{Patterns in Dangerous Speech Across Events}
As a result of right-wing leaders in India strategically conflating right-wing nationalist alignment with patriotism and national security, both social media and mainstream media are popularly used by dangerous users to attribute ``anti-national'' intent to vulnerable minorities, dissenters, and government critics \cite{oommen1986insiders}. This form of othering supplants target identities - in this case, Muslims, Sikhs, and farmers - with those of terror and conspiracy. This context allows us to understand dangerous speech as a weapon in an arsenal of political strategies that rely on posing vulnerable minorities as a threat to the integrity of the nation, in order for political leaders and influencers to both achieve political influence and invoke majoritarian insecurity, victimhood, and polarisation towards ends of political power.

Among semantic choices by dangerous speech users, we find evidence for what can be termed as “popularized synonymisation and replacement” of target groups. Hashtags, keywords and phrases used by political leaders and influencers bring a charge of extremism against target groups, lowering the threshold for violence against them. Within our dataset of tweets, including dangerous speech surrounding the anti-CAA protests and the Tablighi Jamaat incident, the term ``jihadi'' occurs approximately three times more frequently than ``Muslim''. Within tweets published during the farmers’ protests, the term ``khalistani'' occurs upto ten times more frequently than ``Sikh''. Indeed, the popularization of extremist labels has an adverse affect on how minorities are constructed within popular discourse. The trope of ``jihad'', for example, is a central element of othering \cite{chatterjee2012they, kumar2013constructing}. During the first wave of the COVID-19 pandemic, dangerous speech surrounding the Nizamuddin gathering triggered widespread calls for boycotting Muslims across India, leading to Muslims being beaten in several reported incidents for supposedly carrying the virus \cite{prasad2020organization}. Similarly, the synonymisation and replacement of the Sikh farmer identity with the ``khalistani'' label effectively transferred the farmers' protests against agricultural legislation into the secessionist realm. 

Beyond semantics, the dominant narratives that emerge from the dangerous speech-related tweets themselves are heavily laden with disinformative, conspiratorial allegations that depart significantly from ground realities. Disinformation unto itself is a well-established tool of propaganda \cite{martin1982disinformation} that is shaped by, and further exacerbates, political polarisation \cite{soares2021hashtag}, with the authority and trust afforded to political actors, influencers, and opinion leaders playing a critical role in the spread of disinformation \cite{alexandre2021make, benkler2018network, buchanan2019spreading}. A majority of the narratives propagated by the tweets, such as those pertaining to the funding of anti-CAA protests\footnote{https://www.altnews.in/truth-about-sting-claiming-shaheen-bagh-women-were-paid-rs-500-alt-news-newslaundry-joint-investigation/, https://www.altnews.in/media-misreport-lawyers-kapil-sibal-indira-jaising-dushyant-dave-wrongly-linked-with-alleged-pfi-support-for-caa-stir/},  the Tabhligi Jamat’s intent to destabilise India through bioterrorism\footnote{https://www.altnews.in/coronavirus-video-of-an-undertrial-in-mumbai-falsely-viral-as-nizamuddin-markaz-attendee-spitting-at-cop/, https://www.altnews.in/media-misreport-tablighi-jamat-defecating-in-open-after-being-refused-non-veg-food/} and the role of khalistani separatism in the farmers' protests\footnote{https://www.altnews.in/pro-pak-pro-khalistan-chants-at-2019-anti-modi-rally-in-us-revived-amid-farmers-protest/, https://www.altnews.in/2013-image-of-khalistan-supporter-shared-as-recent-farmer-protest/, https://www.altnews.in/old-video-of-pro-khalistan-rally-in-us-linked-with-farmers-protest/, https://www.altnews.in/old-video-from-uk-viral-as-farmers-raising-pro-pak-and-khalistan-slogans-during-protests/} have been debunked as fake news by India-based, IFCN-certified fact-checking agencies. The key role of political leaders and influencers in producing these allegations posits them as sole authors of the truth, thereby affording them authority over biased interpretations of reality that contribute to the expansion of communal rifts within India’s socio-political ecology. High retweet polarity and follower polarity solidify the echo-chambering of such narratives - a finding noted in light of the fact that social media echo chambers also increase political polarisation.

Lastly, while the events that we base our analysis on are distinct, all occurred consecutively within the three-year span of 2019-2021. We note that dangerous speech occurring within each incident gives rise to context, narrative-formation, and legitimization for dangerous speech occurring in subsequent events. This finding is pertinent in light of the fact that propaganda, particularly that which relies on disinformation, is intended to develop antagonistic attitudes that have a long-range effect \cite{martin1982disinformation}. Drivers of dangerous speech frequently linked pro-Muslim and anti-Hindu narratives within CAA and Tabhligi Jamat despite there being little in common among the two events, and also linked Shaheen Bagh, an anti-CAA protest site, to the farmers protests. Further attributions such as anti-national, ``tukde tukde gang" (break India gang), urban naxalism, and anti-Hindu sentiments, developed prior to 2019, were propagated across all events. One such tweet published during the anti-CAA tweets states, \textit{``terrorists maoists naxals separatists khalistanis lutyens leftists isis supporters isi moles dcompany jihadists radical izlamists traitors missioneries rioters goons corrupts pseudo seculars liberals corrupts all above are opposing caa \& nrc''}. This seemingly peculiar phenomenon speaks to the idea that dangerous speech normalises itself over time, builds upon itself, and increasingly solidifies, popularises, and legitimises anti-minority narratives that put vulnerable communities at risk.

\section{Discussion and Future Work} \label{discussion}

Through this paper, we present an overview of dangerous speech networks spearheaded by political influencers on Indian Twitter. Our analysis across three periods of controversy in India precisely identifies dangerous tweets based on both content and context of the messaging using a lexicon based approach followed with manual annotation, and further characterizes dangerous speech users by assigning a danger amplification score computed using a DeGroot model based diffusion algorithm. The scoring mechanism takes into account the currency the speech carries by accounting for influence in the retweet network of the three events. The use of a social learning theory model provides us the scaffolding to look deeper into the characteristics of who engages in dangerous speech, and how. 

\subsection{Dangerous Speech in India}
Among non-dangerous, moderately dangerous, and very dangerous users, moderately dangerous users boast the largest number of followers and the highest percentage of verified accounts, while the very dangerous users have comparable follower counts, the highest friends and favourites counts and the lowest percentage of verified accounts.
Among dangerous speech narratives surrounding the three events, we note a clear pattern of the vilification of minority communities, synonymization and replacement with extremist, anti-national intent, significant propagation of disinformation, and ultimate legitimisation of anti-minority sentiments through long-term discourse-formation. 

Our findings have two critical ramifications. First, dangerous speech users who are considered relatively ambiguous, both through our categorisation as well as in mainstream discourse, as well as those who are highly dangerous, rapidly spread dangerous speech across social media to the end of socio-political adversity and physical violence. Such speech is actively propagated, rewarded and endorsed in increasingly polarised and subsequently radicalised echo chambers. Second, dangerous speech is often subtle in its articulation but impactful in its ultimate consequence. Such speech is communicated in a strategic, coded manner that allows it to pass largely undetected through automated hate speech filters and civility thresholds. Indeed, dangerous speech runs rampant on Twitter, leading to narrative-formation and legitimisation of further dangerous speech and contributing to India’s ongoing political crisis and rising violence against minorities. We also note the key role of mass media personalities and platform celebrities in the propagation of dangerous speech - an indicator of how technologically mediated political propaganda is strategized in the country. Our recognition of these contours is aided by our reflection on offline, localised legacies of polarisation in the region of our study. It is through recognising the baggage of viral dangerous speech in India that we begin to comprehensively understand the logistics at play in framing target groups online, as well as anticipate future impacts of such framing. 

\subsection{Foreboding of Violence}
As noted previously, dangerous speech is an effective propaganda tool as it forms a straightforward path towards the technologically mediated polarisation of vote bases led by political actors, subsumes vulnerable minorities as collateral, and in India’s case, contributes asymmetrically towards the right-wing government’s rise in political power. Perhaps most importantly, dangerous speech also precedes on-ground violence. In February 2020, amidst the ongoing propagation of inflammatory content polarising Muslim and Hindu communities - both online, where Muslims were labelled jihadis and traitors, and offline, where political leaders such as Anurag Thakur, member of parliament from BJP , popularised the chant \textit{``desh ke gaddaron ko, goli maaro saalo ko''} (shoot the traitors of the nation) - India’s capital witnessed the week-long Delhi riots that led to 53 deaths (of which a majority were Muslims), 200+ injuries, and over 2000 arrests and detentions\footnote{https://thewire.in/government/delhi-riots-official-toll-hurt-cases}. During COVID-19, beatings and assault continued to target Muslims for markers of their faith \cite{prasad2020organization}, with hundreds of congregation attendees jailed on criminal charges \cite{rai_2020, shantha_chauhan_2021}. Clashes between farmers and the police at New Delhi’s Red Fort, too, resulted in police violence against protesters. As sit-in protests have continued across seasons, dozens of farmers have died of reasons ranging from hypothermia to suicide, electrocution and traffic accidents \cite{lalwani_2021}.  Further, across the three events, dangerous speech not only served as a forebearer of collective and erratic violence, but its compositions also dominated how violent incidents were rationalised. Apart from physical violence, dangerous speech during the periods we studied also produced justifications for prolonged suffering through dehumanisation and demonization. 

\subsection{Computational Approaches to Dangerous Speech}
The main contribution of this work is showing how computational techniques can be used to pinpoint trends in dangerous speech on social media. The proposed pipeline of identifying, quantifying and characterizing dangerous speech incorporates the hallmark conceptualizations of the dangerous speech framework and enables a large scale computational study of its configurations and spread.

As we characterize users who repeatedly engage in dangerous speech, we find that they are not only more active on Twitter (in terms of statuses and favourites) but also get retweeted more on a group as well as individual level. Indeed, it is the very technological affordances that they can engage more effectively for their purposes, and which in turn enable the spread of their content and positions. Moreover, the audience of these dangerous speech users, in terms of those who retweet them as well as regularly view their content, is more polarized, indicating that the audience of such inflammatory messaging is susceptible. A nuanced analysis in terms of the spectrum of dangerous speech users shows that the moderately dangerous users have the largest audience (in terms of followers) and the highest composition of verified accounts. This shows that when dangerous speech comes from influential individuals, it is delivered in subtle manifestations and innuendo rather than through open attack. 

The composition of the individuals who engage in dangerous speech is of particular concern. Many engage in content creation, as journalists, platform celebrities etc. while there are also right leaning politicians who form the lower echelons of the party. The network centrality analysis suggests that the very dangerous users are not retweeted by the influential nodes in the network, but act as broadcasters, wherein they are able to disseminate their messaging quickly to the rest of the network. The prominent and consistent role, across all the events, of mainstream media accounts as propagators of dangerous speech are of cause for concern, indicative also of the perilous polarization that they pander to.
The observations of our paper further find resonance with research from the social computing community on understanding politicised speech on Twitter. Our reflections on dangerous speech shed light on the performativity of political communications by elite actors that \citet{DBLP:journals/cscw/GonawelaPTVOC18} have discussed, through the realm of coded language. While the authors find that tweets from the Indian Prime Minister do not have the same level of antagonism as other populist leaders under study, our analysis shows that members of his party regularly employ dangerous language instead of direct hate speech, while communicating with their network about out-groups. Viewed in combination with \citet{paakki2021disruptive}'s detailed analysis of trolling strategies that change the track of public conversation, quantified analyses of dangerous speech, such as ours, can further help understand the dominant flows through which communication is mediated on Twitter, as well as the instances when inflammatory language can cause harm.

We hope that our novel approach in quantifying dangerous speech at a large scale can be useful for future computational studies that seek to leverage the dangerous speech framework to identify the usage of explicit and codified references targeting minority groups as well as characterize those who engage in such inflammatory speech acts. We also hope to inspire qualitative studies of the spread of rumours on-ground in times of violence, which can further help us understand the interlacing causational mechanisms at play in the orchestration of violence, especially after the popularization of social media. 

\subsection{Limitations and Future Work}
One of the limitations of our study is that the dangerous speech lexica for each of the events is not exhaustive. While we were able to precisely identify dangerous tweets using our lexicon set, we are not able to identify all the dangerous tweets. Second, we consider only tweets in English or Romanised Hindi in our analysis, and filter out the other tweets. While tweets in Indian scripts are emerging as a rich space for understanding code-mixing and linguistically altered discourse \cite{srivastava2020understanding}, the shared knowledge of Hindi and localised cultural signifiers between all authors allowed us to reflect on the content of the tweets, when Hindi does appear in Roman script. As part of future work we aim to explore additional automated approaches for identifying code-mixed dangerous speech, such as the use of transformer models in \citet{sharif2021nlp,dowlagar2021offlangone}. Our research is a formative step in extending these quantitative methods to expansive corpuses of tweets and accurately capturing the prevalent trends of dangerous speech on Indian Twitter.

\bibliographystyle{ACM-Reference-Format}
\bibliography{ref.bib}

\end{document}